\documentclass[useAMS,usenatbib]{mn2e}
\usepackage{graphicx}
\usepackage{wrapfig}
\makeatletter

\newcommand{\Rmnum}[1]{\expandafter\@slowromancap\romannumeral #1@}
\makeatother

\title[Obscuration in galaxies and SFHs]{Galaxy and Mass Assembly (GAMA): Dust obscuration in galaxies and their recent star formation histories}

\author[D. B.Wijesinghe et al.]
{D. B. Wijesinghe$^{1}$\thanks{E-mail:D.Wijesinghe@physics.usyd.edu.au},
A. M. Hopkins$^{2}$, R. Sharp$^{2}$, M. Gunawardhana$^{1}$, S. Brough$^{2}$,
\newauthor
E. M. Sadler$^{1}$, S. Driver$^{3}$, I. Baldry$^{4}$,  S. Bamford$^{5}$, J. Liske$^{6}$, J. Loveday$^{7}$, P. Norberg$^{8}$,
\newauthor
J. Peacock$^{8}$, C. C. Popescu$^{9}$, R. Tuffs$^{10}$, J. Bland-Hawthorn$^{1}$,  E. Cameron$^{11}$, S. Croom$^{1}$,
\newauthor
C. Frenk$^{12}$, D. Hill$^{3}$, D. H. Jones$^{2}$, E. van Kampen$^{6}$, L. Kelvin$^{3}$, K. Kuijken$^{13}$,
\newauthor
B. Madore$^{14}$, B. Nichol$^{15}$, H. Parkinson$^{8}$, K. A. Pimbblet,$^{16}$, M. Prescott$^{4}$, 
\newauthor
A. S. G. Robotham$^{3}$, M. Seibert$^{13}$, E. Simmat$^{10}$, W. Sutherland$^{17}$, E. Taylor$^{1}$,
\newauthor
D. Thomas$^{12}$ \\ 
$^{1}$Sydney Institute for Astronomy, School of Physics, University of Sydney, NSW 2006, Australia\\
$^{2}$Anglo Australian Observatory, PO Box 296, Epping, NSW 1710, Australia\\
$^{3}$School of Physics \& Astronomy, University of St Andrews, North Haugh, St Andrews, KY16 9SS, UK\\
$^{4}$Astrophysics Research Institute, Liverpool John Moores University, Twelve Quays House, Egerton Wharf, Birkenhead, CH41 1LD, UK\\
$^{5}$Centre for Astronomy and Particle Theory, University of Nottingham, University Park, Nottingham, NG7 2RD, UK\\
$^{6}$European Southern Observatory, Karl-Schwarzschild-Str.~2, 85748, Garching, Germany\\
$^{7}$Astronomy Centre, University of Sussex, Falmer, Brighton BN1 9QH, UK\\
$^{8}$Institute for Astronomy, University of Edinburgh, Royal Observatory, Blackford Hill, Edinburgh, EH9 3HJ, UK\\
$^{9}$Jeremiah Horrocks Institute, University of Central Lancashire, Preston PR1 2HE, UK\\
$^{10}$Max Planck Institute for Nuclear Physics (MPIK), Saupfercheckweg 1, 69117, Heidelberg, Germany\\
$^{11}$Department of Physics, Swiss Federal Institute of Technology (ETH-Z{\" u}rich), 8093 Z{\" u}rich, Switzerland\\
$^{12}$Institute for Computational Cosmology, Department of Physics, Durham University, South Road, Durham, DH1 3LE, UK\\
$^{13}$Leiden University, P.O.~Box 9500, 2300 RA Leiden, The Netherlands\\
$^{14}$Carnegie Institution for Science, 813, Santa Barbara Street, Pasadena, California, 91101\\
$^{15}$Institute of Cosmology and Gravitation (ICG), University of Portsmouth, Dennis Sciama Building, Burnaby Road, Portsmouth PO1 3FX, UK\\
$^{16}$School of Physics, Monash University, Clayton, Victoria 3800, Australia\\
$^{17}$Astronomy Unit, Queen Mary University London, Mile End Rd, London E1 4NS, UK\\}
\begin{document}

\date{Accepted 2010 August 27}

\pagerange{\pageref{firstpage}--\pageref{lastpage}} \pubyear{2009}

\maketitle

\label{firstpage}

\begin{abstract}
We present self-consistent star formation rates derived through pan-spectral analysis
of galaxies drawn from the Galaxy and Mass Assembly (GAMA) survey. We determine the most
appropriate form of dust obscuration correction via application of a range of
extinction laws drawn from the literature as applied to H$\alpha$, [O{\sc ii}] and 
UV luminosities. These corrections are applied to a sample of 31\,508
galaxies from the GAMA survey at $z < 0.35$.
We consider several different obscuration curves, including those of Milky Way, Calzetti (2001) and 
Fischera and Dopita (2005) curves and their effects on the observed luminosities.
At the core of this technique is the observed Balmer decrement, and we provide a prescription to apply optimal 
obscuration corrections using the Balmer decrement.
We carry out an analysis of the star formation history (SFH) using stellar population synthesis tools to
investigate the evolutionary history of our sample of galaxies as well as to understand the effects of variation in the 
Initial Mass Function (IMF) and the effects this has on the evolutionary history of galaxies.
We find that the \citet{FD:05} obscuration curve with an $R_{v}$ value of 4.5 gives the best agreement between 
the different SFR indicators. The 2200\,\AA\ feature needed to be removed from this curve to obtain complete 
consistency between all SFR indicators suggesting that this feature may not be common in the average integrated attenuation of galaxy emission.
We also find that the UV dust obscuration is strongly dependent on the SFR.

\end{abstract}

\begin{keywords}
galaxies: evolution -- galaxies: formation -- galaxies: general -- galaxies: initial mass function
\end{keywords}

\section{Introduction}
The absorption of light by dust in galaxies presents a  major challenge in
measuring accurate star formation rates (SFRs) in galaxies. Understanding star formation is
essential to our understanding of the evolutionary history of the Universe on both
local and global scales. SFRs are commonly derived from H$\alpha$ and ultraviolet (UV)
luminosities, both of which are significantly affected by dust obscuration. The basic approach to correcting for this
relies on having an observed constraint on the level of the attenuation, such as
the Balmer decrement \citep{Stn:79, Cdl:89, Cal:01} or the $\beta$-parameter \citep{Cal:94,MHC:99, AS:00} together with a model for the 
wavelength-dependence of the attenuation \citep{Stn:79,Cdl:89,Cal:01, Pir:04, Tuf:04, FD:05}. 

Dust has the dual effect of not only reducing the overall luminosity observed from a galaxy but also imprinting
wavelength dependant spectral variations \citep{Cal:97a}.
Dust absorbs short-wavelength radiation and re-emits this energy at longer wavelengths.
Dust also plays a significant role in the 
fluid dynamics, chemistry and star formation of galaxies
\citep{Drn:07}. Even though dust plays a crucial role in galaxies,
quantifying its effects and physical properties remains challenging.
Understanding the roles played by inclination angles, optical depths and galaxy
morphologies when coupled with dust has been even more difficult, and only recently have these effects been quantified by \citet{Tuf:04}.

Correcting for the effects of dust attenuation is usually done using semi-empirical models
\citep[see][]{Cdl:89,Byn:94,Cal:01} but there are radiative transfer (RT) models 
developed for attenuation corrections \citep{Xil:97,Xil:98,Xil:99,Wit:92,Tuf:04}.
Even though modelling techniques can give more accurate
attenuation corrections than semi-empirical models, they require more observational information
about the galaxies in order to produce accurate corrections on an object-by-object basis.
However, the RT techniques have proven extremely powerful when utilised to create libraries
of model simulations \citep{Tuf:04} for use in statistical analysis of large statistical samples \citep{Drv:07}.

An important method for providing accurate SFRs is the spectral energy distribution (SED) modelling technique.
There are two main types of SED techniques: template fitting SEDs \citep{Cal:00,Slm:07} and RT SED techniques \citep{Slv:98, Pop:00}.
In particular the accuracy of the RT SED techniques is high, but it tends to be complex and it requires knowledge of the SED over the whole electromagnetic spectrum, from UV to far-infrared (FIR)
and sub-millimeter wavelengths, which is not always available. Due to this complexity the use of RT techniques are less favoured 
than empirical methods for dust correction.

Balmer decrements are commonly used as an indicator of obscuration corrections and are usually applied in conjunction with an extinction law
such as those for the Milky Way (MW) \citep{Nnd:75,Stn:79, Cdl:89}, Small Magellanic Cloud \citep[SMC]{Prv:84, Bch:85}, Large
Magellanic cloud \citep[LMC]{Ftz:86}, that observed for M31 \citep{Bnc:96}, starburst galaxies \citep{Cal:97a,Cal:01} and other galaxies \citep[FD05 hereafter]{FD:05}.
Any extinction law can be used to correct either nebular emission lines or continuum emission. As a consequence of how these curves are derived, though, some curves are preferred over 
others to correct for each type of emission, as the processes that result in these two types of emission are different and it has been shown that nebular lines are usually obscured more than 
the continuum emission \citep{Fan:88, Cal:94, Cal:97b, MK:99}. A physical interpretation for this is provided by \citet{Kel:93} and \citet{Cal:94}
who argue that the ionizing hot (young) stars which produce the nebular lines are found in or close to the dusty molecular clouds from which they were born, while the UV
continuum is a product of older stars that have over time moved away from the dust clouds in which they formed, or have destroyed the dust in situ revealing the cluster of stars.
 
The obscuration curves are also dependent on their environments, in particular the metallicity of the host galaxy \citep{Cdl:89}. 
For instance, the LMC, SMC and MW curves have a large variation in values at the FUV end of the spectrum 
due to the different metallicities of the three galaxies and grain size distribution \citep{Cal:94,Cal:97b} in the interstellar medium.  
These LMC, SMC and MW obscuration curves only take into account the dust between the 
observer and the star but in the case of external galaxies the dust geometry is more complicated. Not only does the dust between the observer and the stars of the galaxy 
need to be taken into account, but back-scattering of light into the line of sight from dust in other regions of the galaxy must also be considered and the dust is never
uniformly distributed. The Calzetti extinction curve attempts to consider all these issues by folding a variety of obscuration effects into a single expression.

\citet{MHC:99} show a relationship between the ratio of far-IR (FIR) and UV fluxes and the UV spectral slope $\beta$
for a sample of starburst galaxies. Because this technique relates the FIR and UV radiation emitted from galaxies it can be a powerful tool in recovering the UV radiation lost 
due to the dust, regardless of the geometry of the dust. The $\beta$ parameter is based on the relation between the gradient in the UV
obscuration curve and the UV wavelengths \citep{MHC:99, Cal:94, AS:00}. However, this method is only recommended for starburst galaxies as 
quiescent galaxies tend to deviate from the total FIR to UV luminosity ratio and UV spectral slope relation seen for starbursts \citep{Kng:04}.
The accuracy of this method is highly dependent on the method employed to measure the UV spectral slope. \citet{Cal:94}
used 10 bands along the observed frame UV continuum spectrum avoiding all large scale features that deviate from the trend of the slope. With large scale surveys such as GALEX this level
of accuracy is unattainable, leading to less accurate measurements of the slope and potentially flawed dust corrections.

Once suitable dust corrections are in hand we can derive accurate SFRs and examine the relationship between galaxy properties and the dust content in a galaxy. 
SFRs are related to the luminosity through a linear scale factor, determined with the assumption of a constant stellar initial mass function (IMF).
These conversion factors will change if the IMF is varied. One recently suggested
possibility is an evolving IMF \citep{Wlk:08a, Wlk:08b, Gun:10} which would lead to an evolving SFR conversion factor. \citet{Pfl:09} proposed an Integrated Galaxy IMF
(IGIMF) that combines the IMFs of all young-star clusters to form a galaxy-wide IMF. This was developed to account for the inconsistencies in current IMFs which are based on isolated 
stellar clusters and then applied on galaxy wide scales. One of the inconsistencies that the IGIMF accounts for is the discrepancy between FUV and H$\alpha$ derived SFRs at 
low SFRs \citep{Sul:00, Lee:09}.

We compare and contrast SFRs corrected for dust using different obscuration correction methods, 
identifying an optimum approach, and then use that in a preliminary investigation of
SFR histories of galaxies. The SFRs are compared with theoretical evolutionary
paths to better understand the star formation history of the galaxies in our sample. 

In $\S$2 we present details of the data used.
In $\S$3 we present a prescription for the derivation and the application of the obscuration corrections. $\S$4 compares H$\alpha$, FUV and 
[O{\sc ii}] derived SFRs. $\S$5 examines the evolution of our SFRs by comparing these SFRs to evolutionary synthesis models. 
We also compare the evolutionary paths of the SFRs with the predictions of the IGIMF theory.
$\S$6 is the summary and conclusion of our findings.  The assumed cosmological parameters are:

$H_{0}=$70\,km\,s$^{-1}$\,Mpc$^{-1}$, $\Omega_{M}=0.3$ and $\Omega_{\Lambda}=0.7$.
All magnitudes are in the AB system.

\section{Data}
We use data from the Galaxy and Mass Assembly (GAMA) survey
\citep{Drv:09}. GAMA is a multi-band imaging and spectroscopic survey covering $\approx$144 square degrees of sky in three
$12^{\circ} \times 4^{\circ}$ regions \citep{Rob:09, Bld:09}.
The spectroscopy comes from the AAOmega spectrograph \citep{Shp:06} at the Anglo Australian Telescope (AAT).

The UV data was obtained from the GALEX-GAMA survey \citep{Seb:10}.
A signal-to-noise ratio cut of 2.5 was applied to the UV data.
There are 111521 GALEX sources within the GAMA regions of which
110443 matching sources were found using a matching radius of $6^{\prime \prime}$. Such a large matching radius is is necessary due to
the increased positional uncertainty inherent in the larger GALEX PSF.
SDSS and spectral line data were matched to GAMA sources using a matching radius of $2.5^{\prime \prime}$ resulting in a final sample of
47269 objects. After removing AGN (using the prescription of Kewley, 2006) and sources without emission lines, $r$-band magnitude and redshift data the final sample 
size contained 31508 galaxies. All remaining galaxies contain both H$\alpha$ and H$\beta$ fluxes. 
The final sample has a GALEX FUV magnitude range of $14.7 < m_{FUV} < 25.3$ and a redshift limit of $z < 0.350$ due to the requirement for H$\alpha$ being in the observed spectral range.

\section{Dust Obscuration Corrections}
Correcting galaxy emission for dust requires a case specific approach, that is nebular and continuum emission from galaxies require two different treatments.
At the core of these obscuration corrections are the obscuration curves and the Balmer decrement. 
We apply several different obscuration curves to examine their effects on continuum (UV) and nebular (H$\alpha$ and [O{\sc ii}]) emission.

\subsection{Continuum emission corrections}

Applying obscuration corrections for continuum emission requires two parts, the reddening suffered by the stellar continuum ($E(B-V)_{cont}$) and an applicable obscuration curve ($k(\lambda)$).
We use the standard form
\begin{equation}
L_{i} = L_{o}10^{0.4E(B-V)_{cont}k(\lambda)}.
\label{eq:Cal_law}
\end{equation}
to apply our dust corrections. $L_{i}$ and $L_{o}$ are the intrinsic and observed galaxy luminosities respectively. To obtain $k(\lambda)$ for the UV stellar continuum, 
the \citet{Cal:01} obscuration curve and the FD05
obscuration curves can be applied. When using the \citet{Cal:01} curve $k(\lambda)$ must be divided by 0.44.
The FD05 curves can also be applied to correct for the obscuration of nebular emission.
The UV luminosities were derived using the UV fluxes from the GALEX survey and the redshifts obtained from the GAMA survey.

$E(B-V)_{cont}$ is derived from the reddening in the ionized gas, $E(B-V)_{gas}$.
$E(B-V)_{cont} = 0.44E(B-V)_{gas}$ \citep{Cal:01} and
\begin{equation}
E(B-V)_{gas} = \frac{\log\left(\frac{f_{H\alpha}}{f_{H\beta}}/2.86\right)}{0.4[k(H\beta)-k(H\alpha)]},
\end{equation}
where $k(H\beta)$ and $k(H\alpha)$ are the obscurations of the nebular emission at the H$\alpha$ and H$\beta$ wavelengths 
derived from a MW obscuration curve \citep{Stn:79,Cdl:89} or a theoretically modelled curve (FD05) as in this paper.
$f_{H\alpha}$ and $f_{H\beta}$ are the stellar absorption corrected but not dust corrected H$\alpha$ and H$\beta$ fluxes.

$f_{H\alpha}$ and $f_{H\beta}$ were corrected for stellar absorption using the formalism outlined in \citet{Hpk:03} stated below.
\begin{equation}
f_s = \left[\frac{EW + EW_{c}}{EW}\right]f_{o}
\end{equation}
where $f_{o}$ and $f_{s}$ are the observed and stellar absorption corrected fluxes respectively. EW is the equivalent width of the line being corrected and $EW_{c}$ is the
correction for stellar absorption taken to be 0.7\,\AA\, which has been shown to be a reasonable assumption \citep{Gun:10}.

Assuming a Case-B recombination with a density of 100\,cm$^{-3}$ and a temperature of 10\,000\,K, the predicted ratio of $f_{H\alpha}$ to $f_{H\beta}$ is 2.86 \citep{Ost:89}.
All Balmer decrements below 2.86 were set equal to 2.86 as suggested by \citet{Kwl:06}.
We applied k-corrections to the observed GALEX UV magnitudes, using {\sc kcorrect.v4.1.4} \citep{Blt:03} to infer the rest frame magnitude in each GALEX band, with effective wavelengths 
of 1528\,\AA\ and 2271\,\AA\ ,
before the dust correction process.

It is important to note that correcting continuum emission for dust using this technique requires both MW and continuum obscuration curves. MW obscuration curves 
such as \citet{Stn:79,Cdl:89} are required for the derivation of $E(B-V)_{cont}$ while a continuum obscuration curve such as \citet{Cal:01} provides $k(\lambda)$.

A theoretically modelled curve such as FD05 is applicable to both the nebular and continuum components.
The reason for this is that unlike the other extinction correction curves, these curves were derived from dust models that do not require assumptions about
the effects of the emission sources of the galaxy. The model is based on the inferred physical properties of the turbulent density structure (FD05).
The FD05 curves are described for a range of $R_{v}$ values and we examine the effects of different $R_{v}$ values. $R_{v} = A_{V} / E(B-V)_{gas}$, where $A_{V}$ is the
extinction suffered in the rest frame $V$-band.

\subsection{Nebular emission-line corrections}
To correct H$\alpha$ emission for dust, the H$\alpha$ luminosity needs to be calculated first.
H$\alpha$ luminosities corrected for stellar absorption, taking into account aperture corrections, can be obtained using the formalism outlined in \citet{Hpk:03},
\begin{equation}
L_{H\alpha} = (EW_{H\alpha} + EW_{c})10^{-0.4(M_{r}-34.10)}\frac{3\times10^{18}}{[6564.61(1+z)]^{2}}.
\end{equation}
$M_r$ is the $k$-corrected absolute $r$-band AB magnitude. The last factor converts units from W Hz$^{-1}$ to W\,\AA$^{-1}$ and 6564.61\,\AA\ is the 
vacuum wavelength of H$\alpha$. 

Aperture corrections account for the fact that only a limited amount of emission from a galaxy is detected through the $2^{\prime \prime}$ diameter AAOmega fiber.
The aperture correction implicitly assumes that the emission measured through the fiber is characteristic of the
whole galaxy and that the star formation is traced by the $r$-band continuum emission \citep{Hpk:03}.

Obscuration corrections are then applied to the H$\alpha$ luminosity as follows,
\begin{equation}
L_{i} = L_{H\alpha}\left(\frac{f_{H\alpha}/f_{H\beta}}{2.86}\right)^{2.942}.
\end{equation}
The exponent in the above equation is derived from the FD05 ($R_{v}=4.5$) extinction curve and will vary according to the (MW) obscuration curve.
This form of obscuration correction has the same end result as
\begin{equation}
L_{i} = L_{H\alpha}10^{0.4E(B-V)_{gas}k(\lambda)}.
\end{equation}
In this case $k(\lambda)$ can be derived from a MW obscuration curve such as \citet{Stn:79,Cdl:89} or a theoretically modelled curve such as FD05.

An analogous method can be used to derive [O{\sc ii}] luminosities
but the stellar absorption correction ($EW_{c}$) is not required. $M_{r}$ is replaced by $M_{u}$ and the H$\alpha$ wavelength is
replaced by the [O{\sc ii}] doublet effective wavelength, 3728.30\,\AA\ \citep{Hpk:03}. As in the case of H$\alpha$, aperture corrections are 
included but obscuration corrections are not.
\begin{equation}
L_{\textrm{[O{\sc ii}]}} = EW_{\textrm{[O{\sc ii}]}}10^{-0.4(M_{u}-34.10)}\frac{3\times10^{18}}{[3728.30(1+z)]^{2}}
\end{equation}

 [O{\sc ii}] emission can be used as an SFR estimator, particularly at high redshifts \citep{Gil:10}.
The [O{\sc ii}] emission must be corrected for obscuration
based on the obscuration at the wavelength of H$\alpha$ so that the SFR calibration derived for H$\alpha$ luminosities (calibrated to [O{\sc ii}]) can be used \citep{Hpk:03,Ken:92,Ken:98}.

\subsection{$\beta$ Parameter}

The $\beta$ parameter \citep{MHC:99, AS:00} has been proposed in order to provide obscuration corrections for starburst galaxies when the Balmer decrement is not available.
It is derived from the relationship between the UV spectral slope and UV to FIR flux ratio.

The $\beta$ parameter is the UV spectral slope determined from a power-law fit to the UV continuum of the form: 
\begin{equation}
f_{\lambda} \propto \lambda^{\beta},
\end{equation}
where $f_{\lambda}$ is the flux density per wavelength interval and $\lambda$ is the central rest wavelength \citep{MHC:99}.
The relationship between the $\beta$ parameter and the ratio of far-infrared and UV fluxes is used to account for dust in galaxies.
\citet{Cal:94} recommend observing the UV continuum bands that avoid irregularities
along the UV spectral slope that would alter the gradient of the slope that is measured.
In our case only 2 UV bands are available, those of far UV (FUV) and near UV (NUV) from the GALEX satellite that span 1400\,\AA\ - 1800\,\AA\ and 1800\,\AA\ - 2800\,\AA\ 
with effective wavelengths $\lambda_{\rm FUV}=1528\,$\AA\ and 2271\,\AA\ respectively.

Due to the poor sampling of the underlying UV spectrum by the broad GALEX UV bands, and
significant contamination by local structure across these bands the accuracy with which the UV
slope can be determined is compromised.
Kong et al. (2004) state that this definition of $\beta$  does not apply to galaxies for which only multi-band ultraviolet imaging is available, as in the case with 
GALEX MIS Survey. Instead Kong et al. (2004) recommend the definition,
\begin{equation}
\beta=\frac{log{{\bar{f}}_{FUV}} - log{{\bar{f}}_{NUV}}}{{log{\lambda}}_{FUV} - {log{\lambda}}_{NUV}}
\end{equation}
where ${\lambda}_{FUV}$ = 1528 and ${\lambda}_{NUV}$ = 2277 are the effective wavelengths of the far and near UV filters of the GALEX satellite.
${\bar{f}}_{FUV}$ and ${\bar{f}}_{NUV}$ are the mean flux densities per unit wavelength through these filters. We aim to compare the the dust corrections from this technique with corrections made
using the Balmer decrement. This would allow us to observe the effect that the $\beta$ parameter will have on dust corrections in our sample of largely quiescent galaxies.

The UV luminosities are corrected for starburst galaxies using the relation $A_{1600} = 4.43 + 1.99\beta$ from \citet{MHC:99}, where $A_{1600}$ is the attenuation at 1600\,\AA\,:
\begin{equation}
L_{i} = 10^{0.4(4.43+1.99\beta)}L_{o}.
\end{equation}
where $L_{i}$ and $L_{o}$ are the FUV dust corrected and uncorrected luminosities.

We apply this formalism to our sample of largely quiescent galaxies to test the extent to which $\beta$ and hence the corrected luminosities and SFRs are overestimated as 
Kong et al. (2004) have suggested.

\section{Comparison of Star Formation Rates}

Star formation rates can be calculated from scaling factors applied to the luminosities. The scaling factors are dependent on the IMF used \citep{Slpt:55,Ken:83}. We use a \citet{BG:03} IMF 
with a mass range of $0.1 M_{\sun}$ to $120 M_{\sun}$. The effects of applying other IMFs \citep{Slpt:55, Ken:83, Scl:98} are explored in \S\,4.  
For H$\alpha$, FUV and NUV luminosities the SFR conversion factors were calculated using a population synthesis model \citep[][PEGASE.2]{FR:97}.

The [O{\sc ii}] SFR conversion factor was derived from the flux ratio of [O{\sc ii}] and H$\alpha$, where $F_{\textrm{[O{\sc ii}]}}/F_{H\alpha} = 0.23$ \citep{Hpk:03}
and the H$\alpha$ SFR conversion factor \citep{Ken:92}. A stellar absorption correction of 0.7\,\AA\\ \citep{Gun:10} is applied before this ratio is obtained. 
The luminosity to SFR calibration for [O{\sc ii}] can be calculated in this way because
the SFRs come from using the [O{\sc ii}] emission as a proxy for the H$\alpha$ emission.
SFRs derived from [O{\sc ii}] luminosities are based on the fact that there is a good correlation between \emph{observed} [O{\sc ii}] and H$\alpha$ line fluxes \citep{Ken:92,Ken:98,Hpk:03}.

For FUV, NUV, H$\alpha$ and [O{\sc ii}] luminosities, SFRs can be calculated using linear scale factors, derived using PEGASE, as
follows,
\begin{equation}
SFR_{FUV}(M_{\sun}\,yr^{-1}) = \frac{L_{FUV}}{1.64 \times 10^{21} \textrm{W Hz}^{-1}}
\end{equation}
\begin{equation}
SFR_{NUV}(M_{\sun}\,yr^{-1}) = \frac{L_{NUV}}{1.56 \times 10^{21} \textrm{W Hz}^{-1}}
\end{equation}
\begin{equation}
\textrm{SFR}_{\textrm{[O{\sc ii}]}}(M_{\sun}\,yr^{-1}) =  \frac{L_{\textrm{[O{\sc ii}]}}}{7.967 \times 10^{33} \textrm{W}}
\end{equation}
\begin{equation}
\textrm{SFR}_{\textrm{H}\alpha}(M_{\sun}\,yr^{-1}) = \frac{L_{H\alpha}}{3.464 \times 10^{34} \textrm{W}}
\end{equation}

These SFR conversion factors are consistent with those of \citet{Ken:98} for H$\alpha$, \citet{Hpk:03} for [O{\sc ii}] and \citet{Mad:98} for UV, when taking into account the choice of IMF 
\citep{BG:03}.

\begin{figure*}
\centerline{\includegraphics[width=62mm, height=60mm]{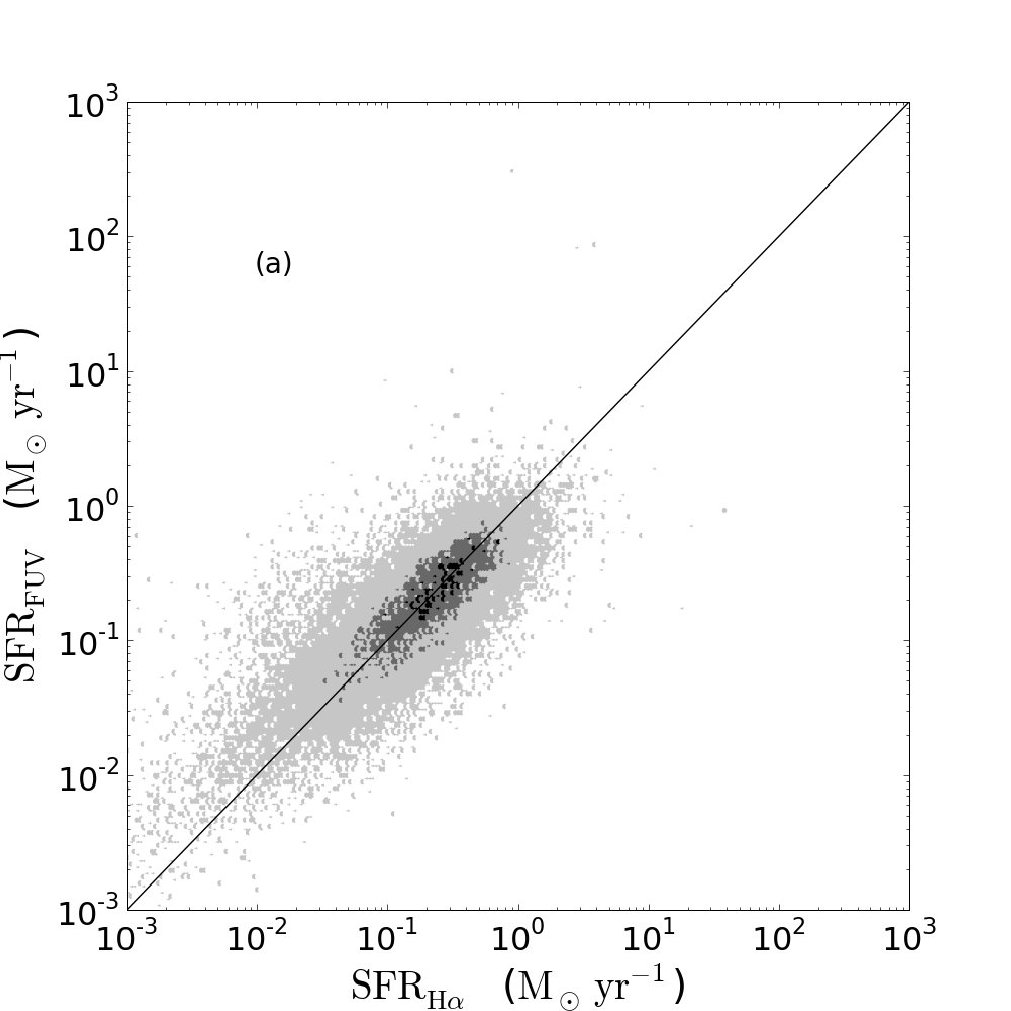}
\includegraphics[width=62mm, height=60mm]{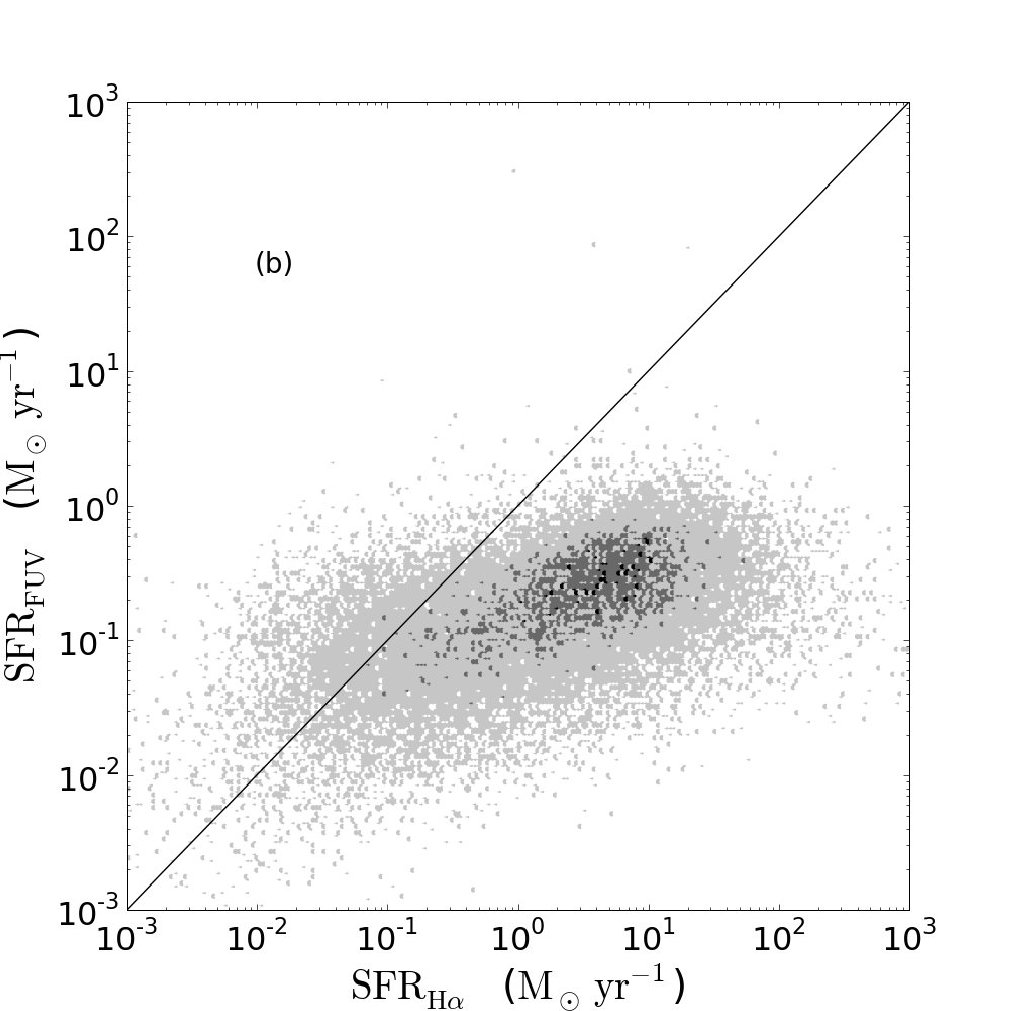}}
\caption{SFR derived from FUV luminosity as a function of SFR derived from H$\alpha$ luminosity. All galaxies in our sample are shown, with AGNs excluded.
(a) Both SFR indicators without obscuration corrections, and (b) only the H$\alpha$ derived SFRs are corrected for dust obscuration using a FD05 curve with
an $R_{v}$ value of 4.5. The axis ranges in the above panels, and subsequent figures, are the same for ease of comparison. 
The strong luminosity dependence of the obscuration corrections is clear (Hopkins et al. 2001, 2003; Afonso et al. 2003).}
\label{fig:sfr_ha_corr_uncorr_fuv_uncorr}
\end{figure*}

Figure~\ref{fig:sfr_ha_corr_uncorr_fuv_uncorr}\,a shows the FUV SFRs as a function of H$\alpha$ SFRs with both indicators uncorrected for dust obscuration, and 
Figure~\ref{fig:sfr_ha_corr_uncorr_fuv_uncorr}\,b shows the same but with only the H$\alpha$ SFRs corrected
for dust obscuration. This shows the effect that obscuration corrections have on the luminosity and how the observed and intrinsic luminosities compare. 
In Figure~\ref{fig:sfr_ha_corr_uncorr_fuv_uncorr} when dust corrections are applied (in this case only to the H$\alpha$), the SFR values of
high SFR objects increase by $\approx$2 orders of magnitude. At low SFRs the distribution remains close to the one-to-one line. 
From similar analyses of uncorrected FUV and H$\alpha$ SFRs by \citet{Sul:00} and \citet{Lee:09}, it is clear that our data follow
similar trends. Even at high SFR, it is clear that our SFRs are a continuation of the same trends observed in \citet{Sul:00} at lower SFRs. 

\begin{figure*}
\centerline{\includegraphics[width=58mm, height=58mm]{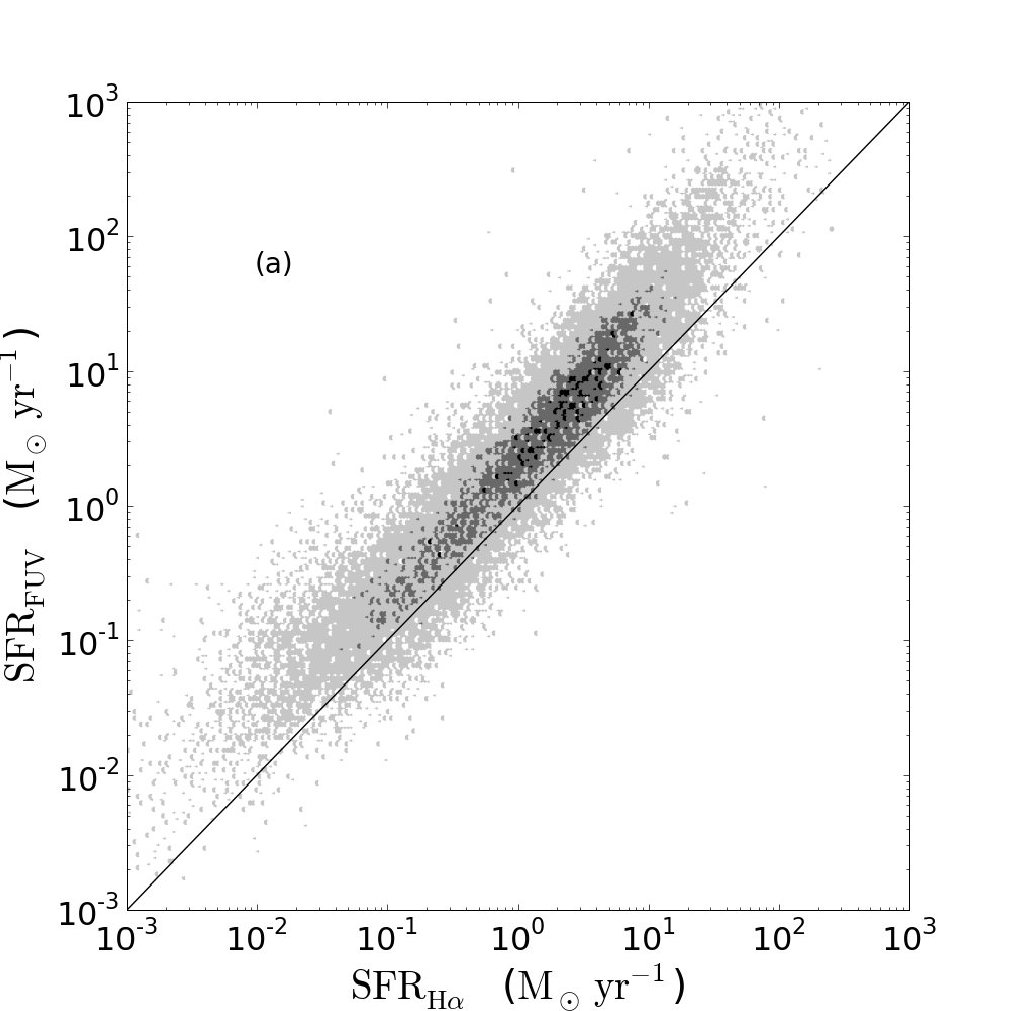}
\includegraphics[width=58mm, height=58mm]{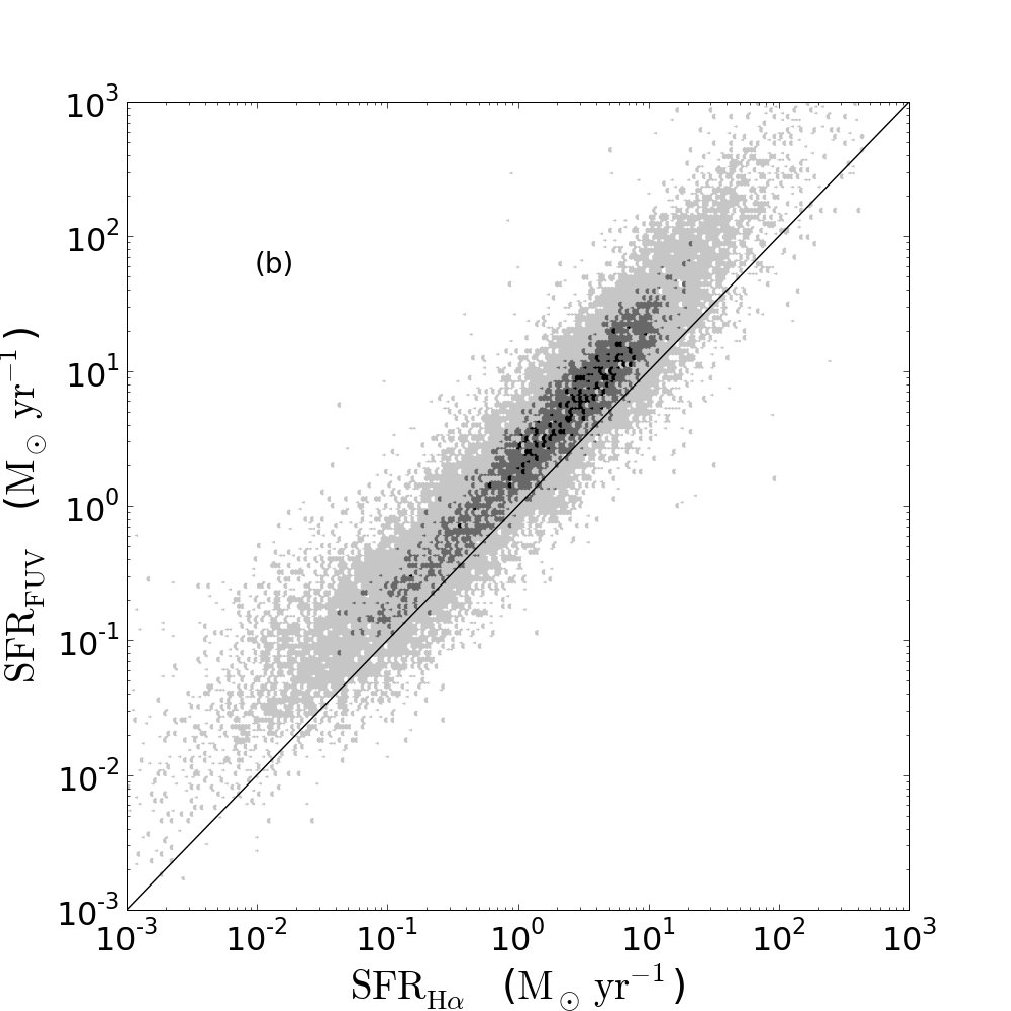}
\includegraphics[width=58mm, height=58mm]{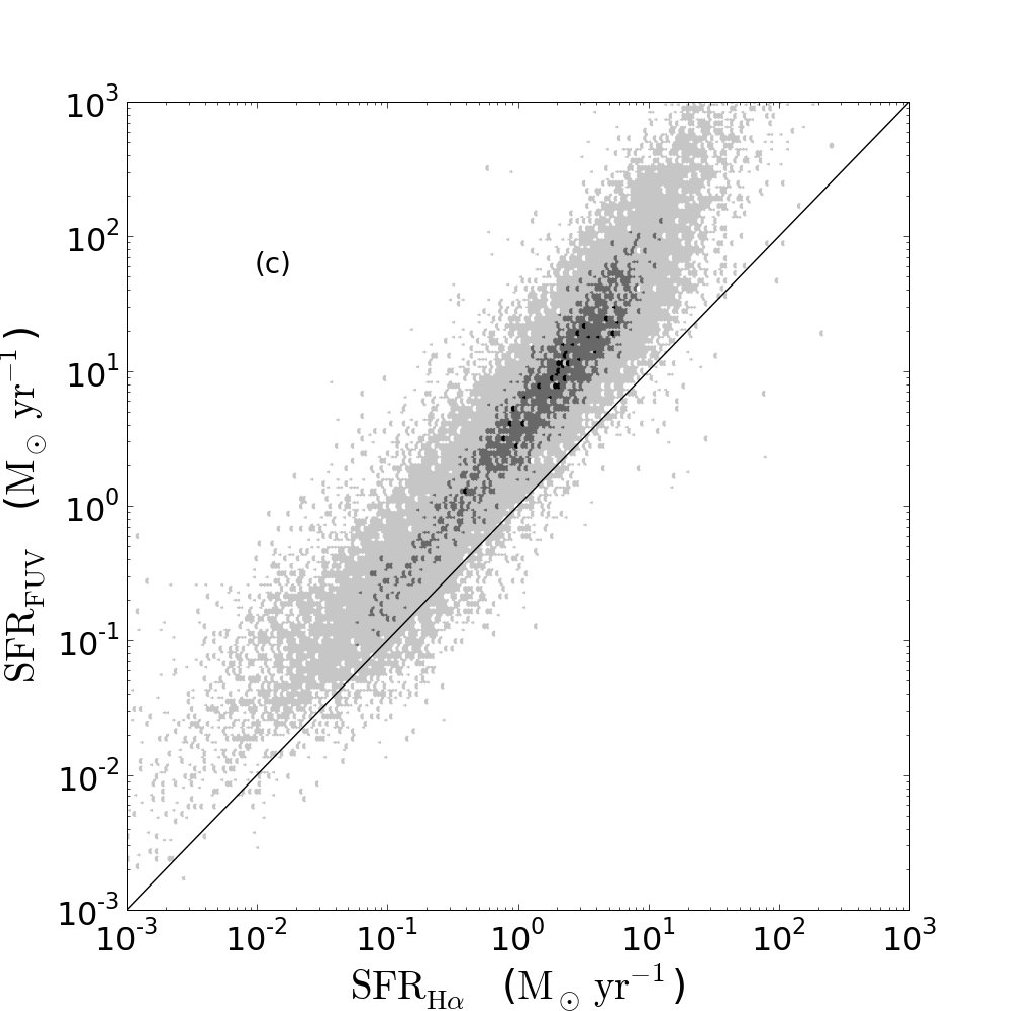}}
\centerline{\includegraphics[width=58mm, height=58mm]{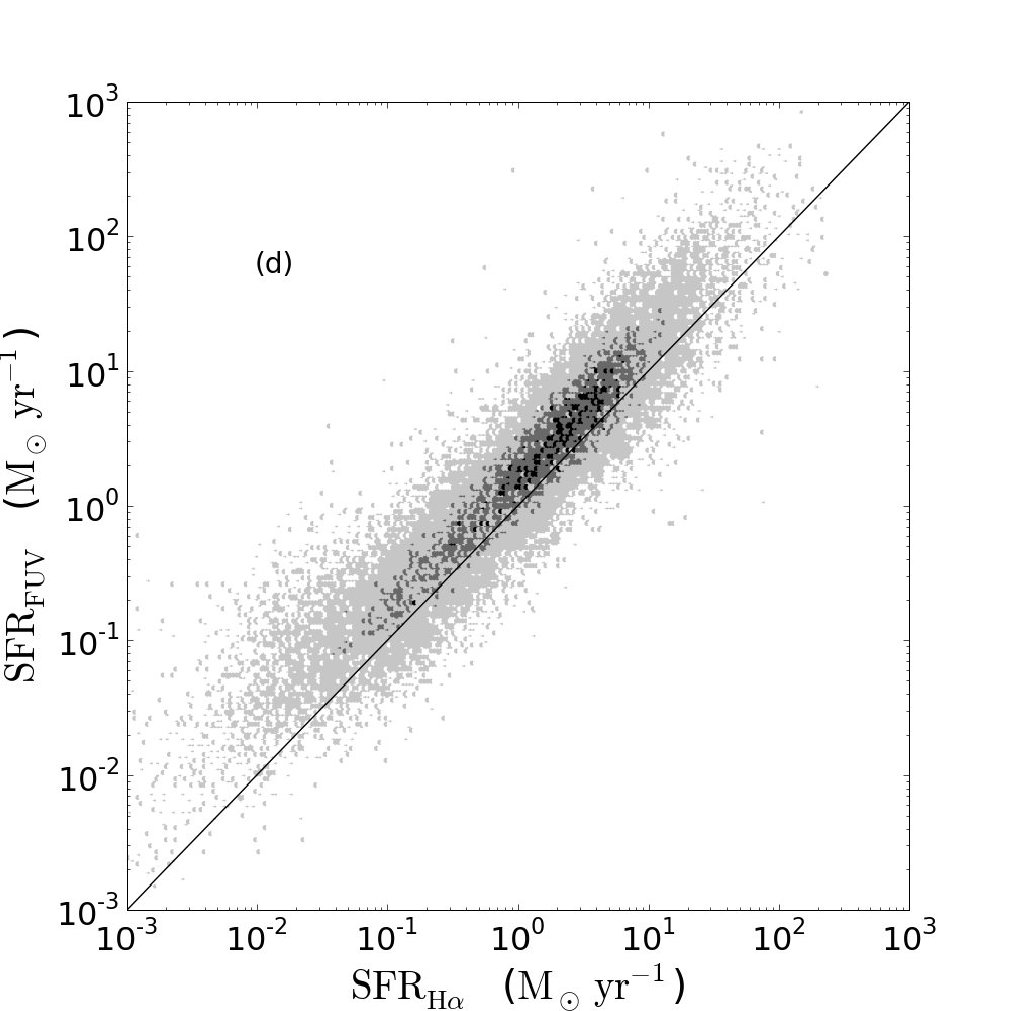}
\includegraphics[width=58mm, height=58mm]{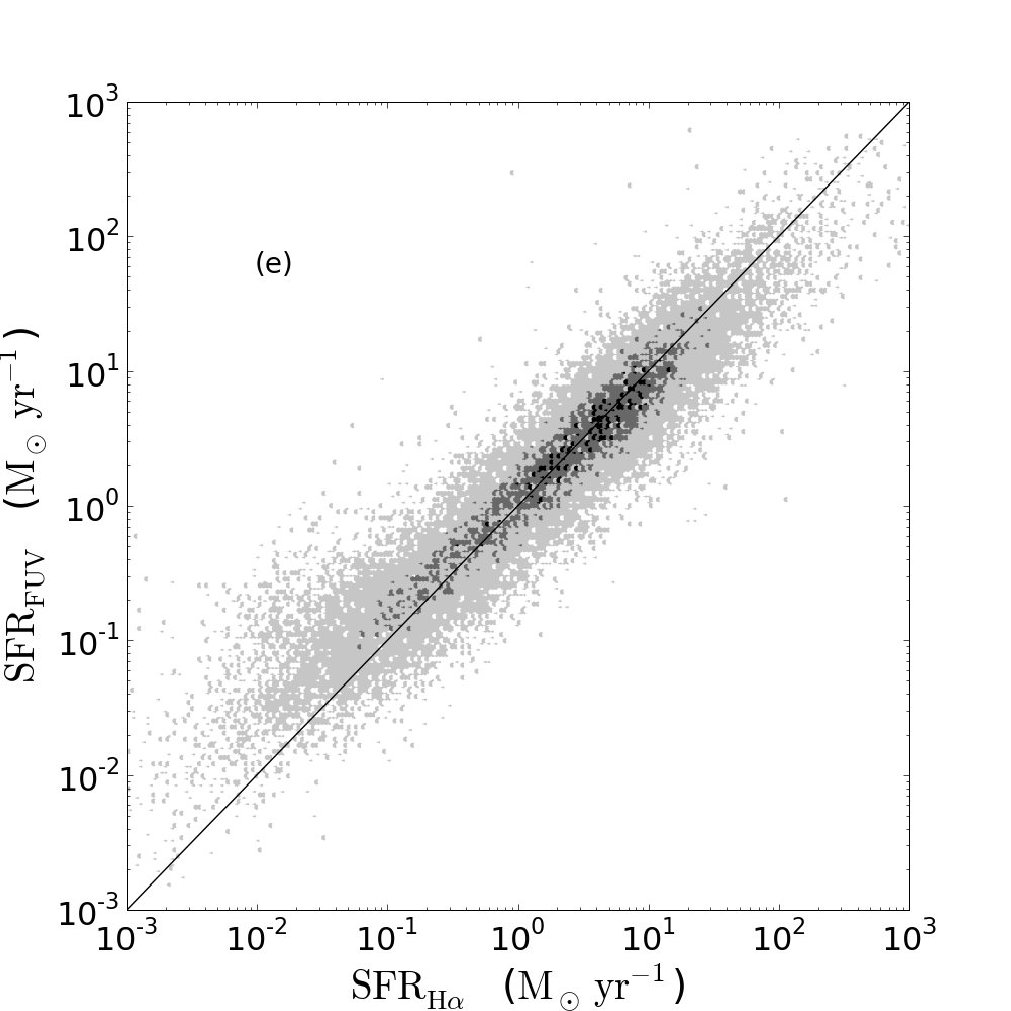}
\includegraphics[width=58mm, height=58mm]{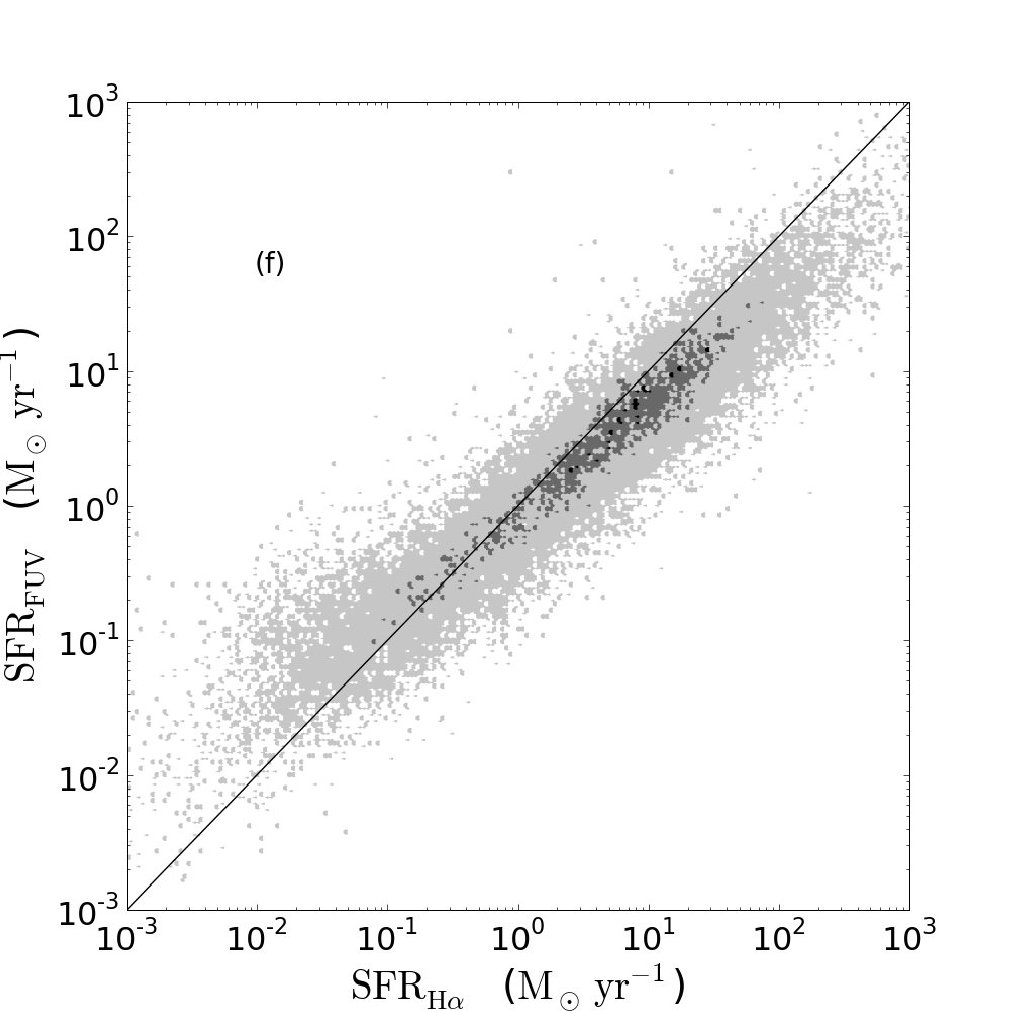}}
\caption{FUV SFRs as a function of H$\alpha$ SFRs, dust-corrected with a variety of obscuration corrections. Both H$\alpha$ and FUV luminosities are corrected with consistent
obscuration curves as described in the text. (a) A Cardelli et al. (1989) extinction curve is used to correct both H$\alpha$ and FUV luminosities. (b) A Calzetti (2001) curve corrects 
both H$\alpha$ and FUV luminosities. (c) Cardelli et al. (1989) and Calzetti (2001) curves correct for the  nebular and continuum emission respectively. (d), (e) and (f) show
SFR values corrected using FD05 extinction curves with $R_{v}$ values of 3.5, 4.5 and 5.5 respectively for both luminosities. The best agreement between H$\alpha$ and FUV derived SFRs is 
the FD05 extinction curve with an $R_{v}$ value of 4.5, panel (e).}
\label{fig:sfr_ha_fuv_same_curve}
\end{figure*}

\begin{figure*}
\centerline{\includegraphics[width=58mm, height=58mm]{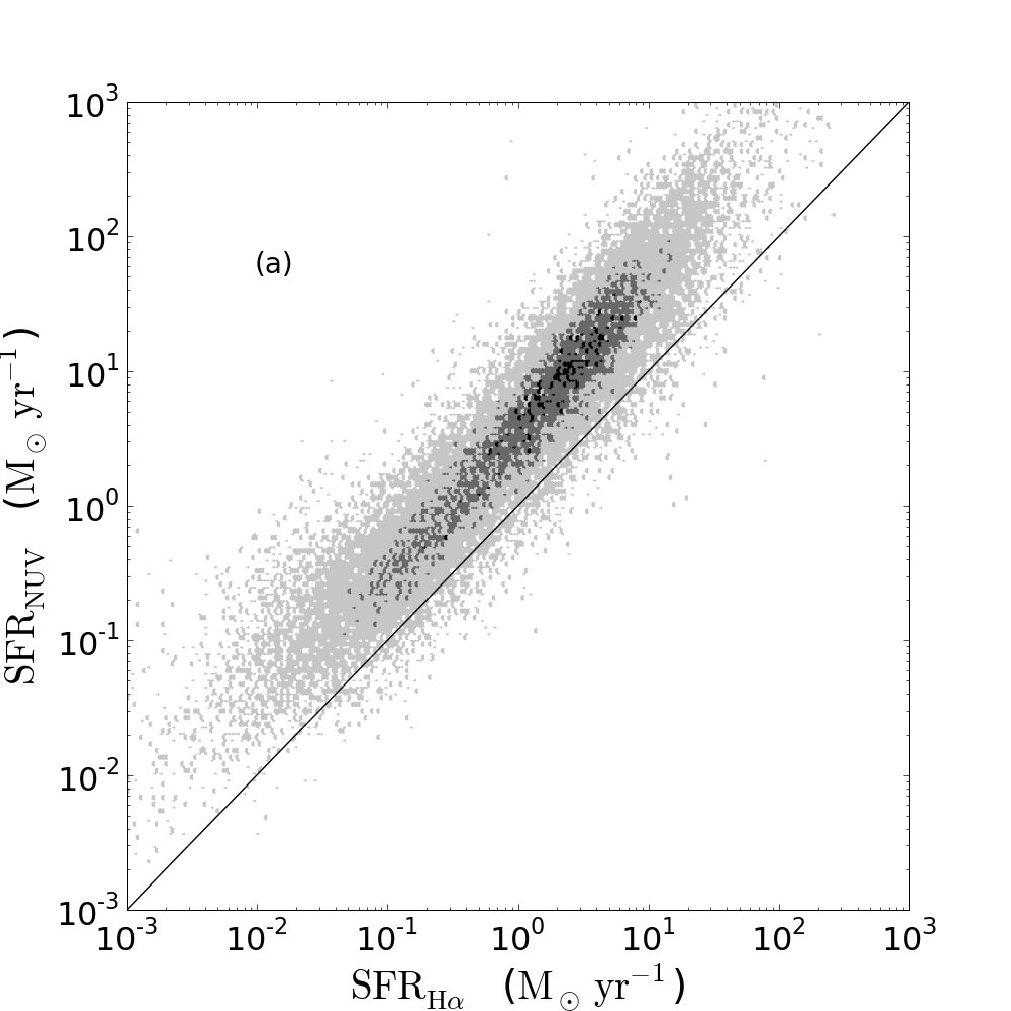}
\includegraphics[width=58mm, height=58mm]{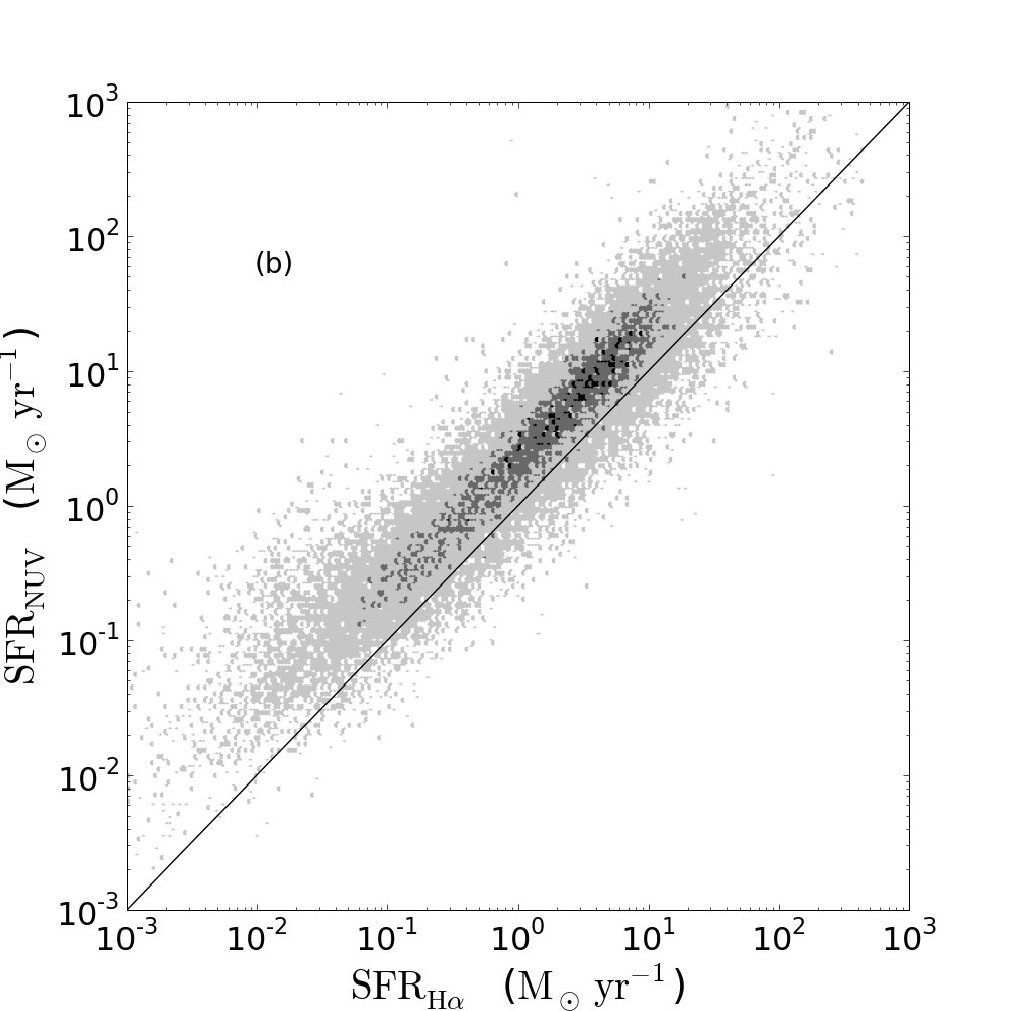}
\includegraphics[width=58mm, height=58mm]{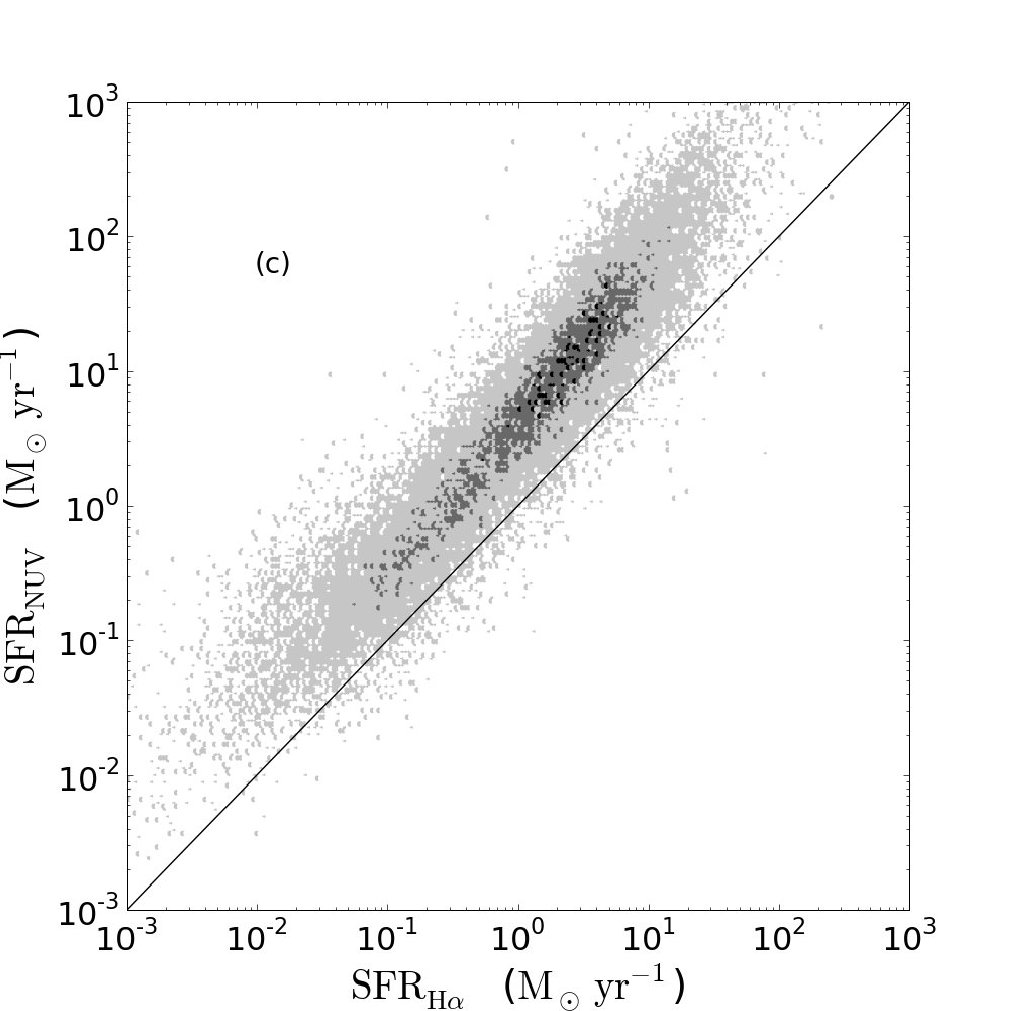}}
\centerline{\includegraphics[width=58mm, height=58mm]{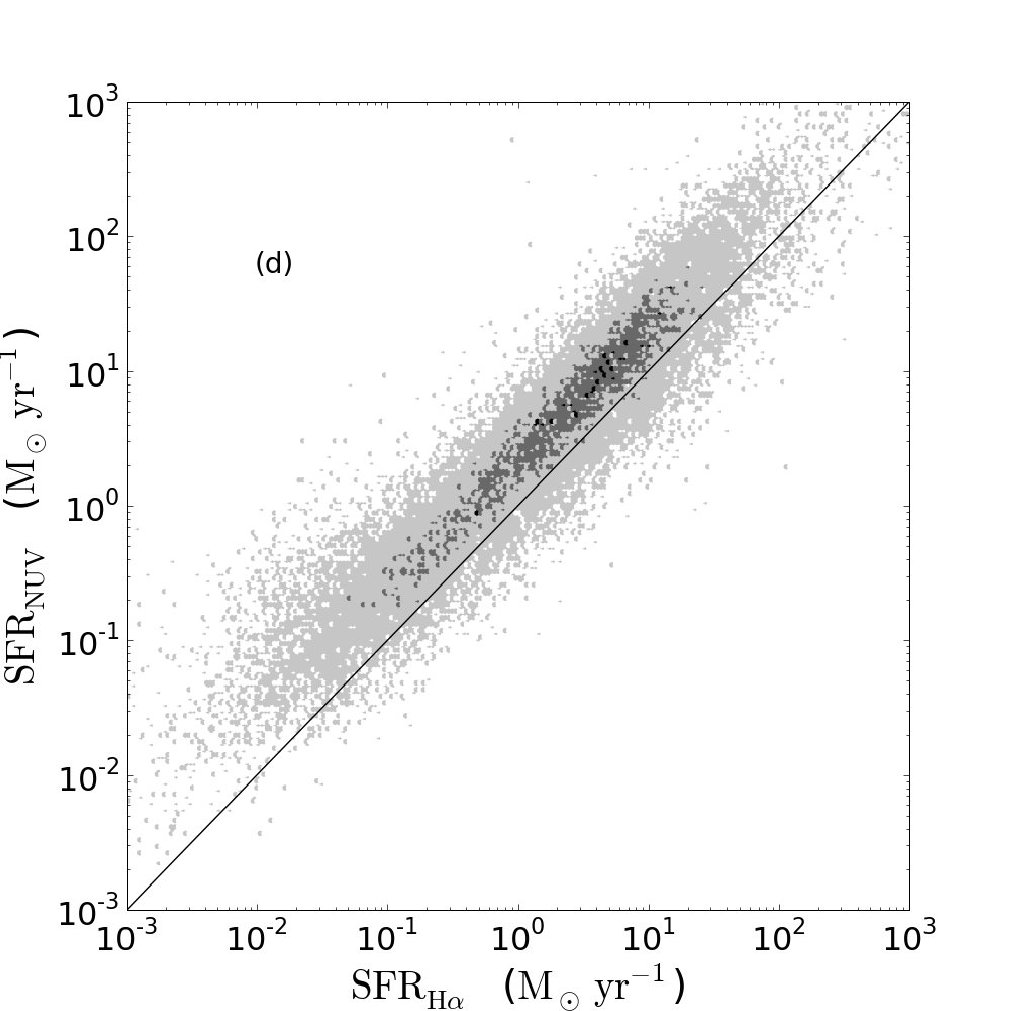}
\includegraphics[width=58mm, height=58mm]{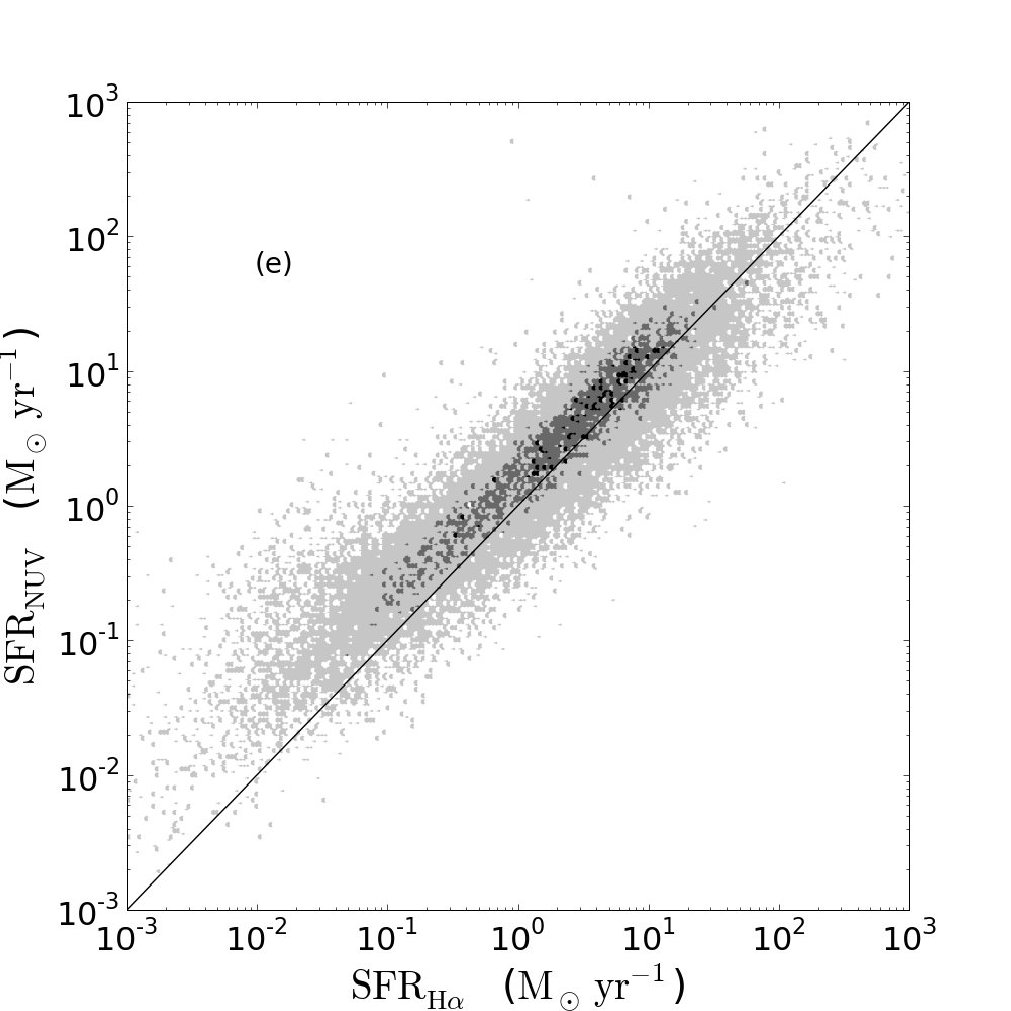}
\includegraphics[width=58mm, height=58mm]{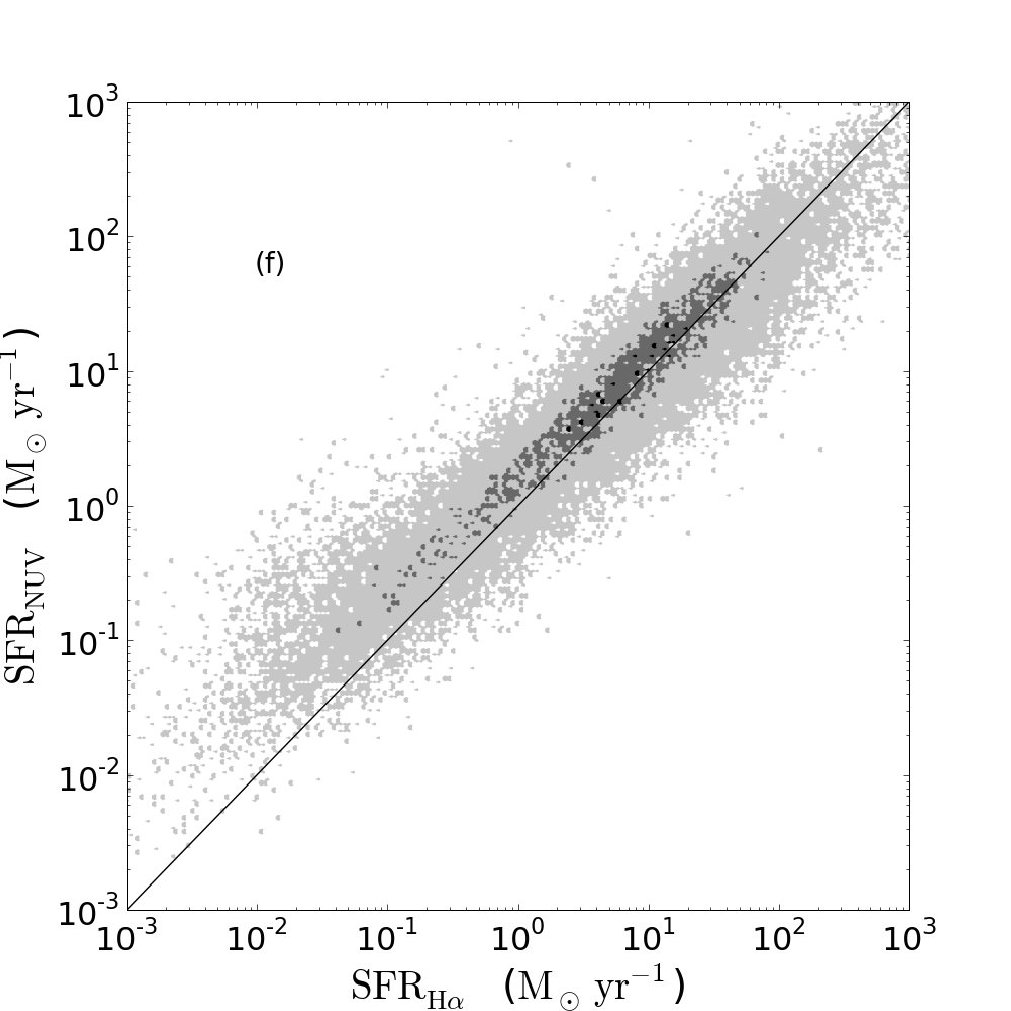}}
\caption{NUV SFRs as a function of H$\alpha$ SFRs dust corrected with a variety of obscuration corrections. Both H$\alpha$ and NUV luminosities are corrected with
consistent obscuration curves. (a) A \citet{Cdl:89} curve was used to correct both H$\alpha$ and NUV luminosities. 
(b) A Calzetti (2001) curve corrects for both H$\alpha$ and NUV luminosities. (c) Cardelli et al. (1989) and Calzetti (2001) curves correct for the  nebular and continuum 
emission respectively. 
(d) A FD05 curve with an $R_{v}$ value of 4.5 and with the 2200\,\AA\ feature is used to correct both H$\alpha$ and NUV. 
(e) A FD05 curve with an $R_{v}$ value of 4.5 and without the 2200\,\AA\ feature is used to correct both H$\alpha$ and NUV. 
(f) A FD05 curve with an $R_{v}$ value of 6 and with the 2200\,\AA\ feature is used to correct both H$\alpha$ and NUV. 
The best agreement between H$\alpha$ and NUV derived SFRs is when the FD05 extinction curve with an $R_{v}$ value
of 6, panel (f), and 4.5 without the 2200 $\textrm{\AA}$ feature, panel (e), are used. This result is at odds with that of Figure (2) as the extinction curves are not
consistent between FUV and NUV luminosities unless we remove the 2200 $\textrm{\AA}$ feature.}
\label{fig:sfr_ha_nuv_same_curve}
\end{figure*}

\begin{figure*}
\centerline{\includegraphics[width=62mm, height=60mm]{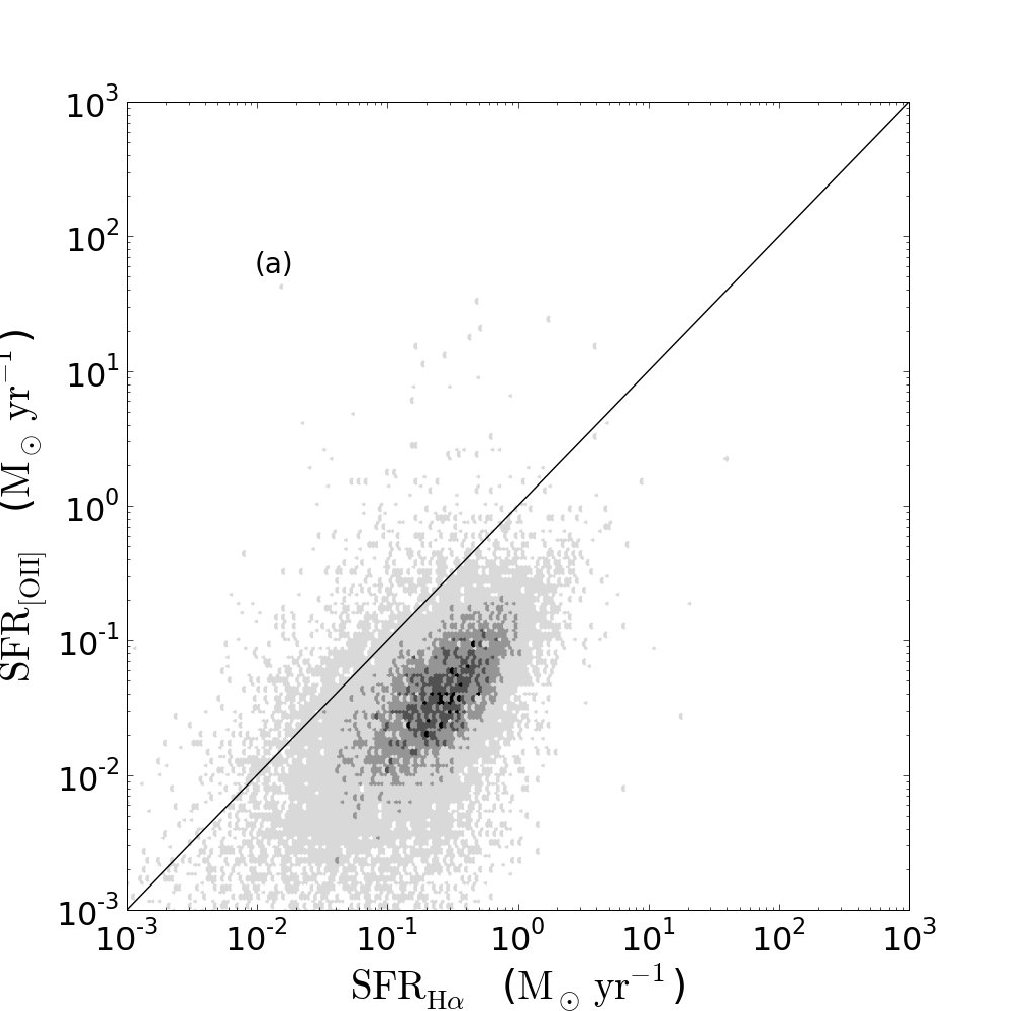}
\includegraphics[width=62mm, height=60mm]{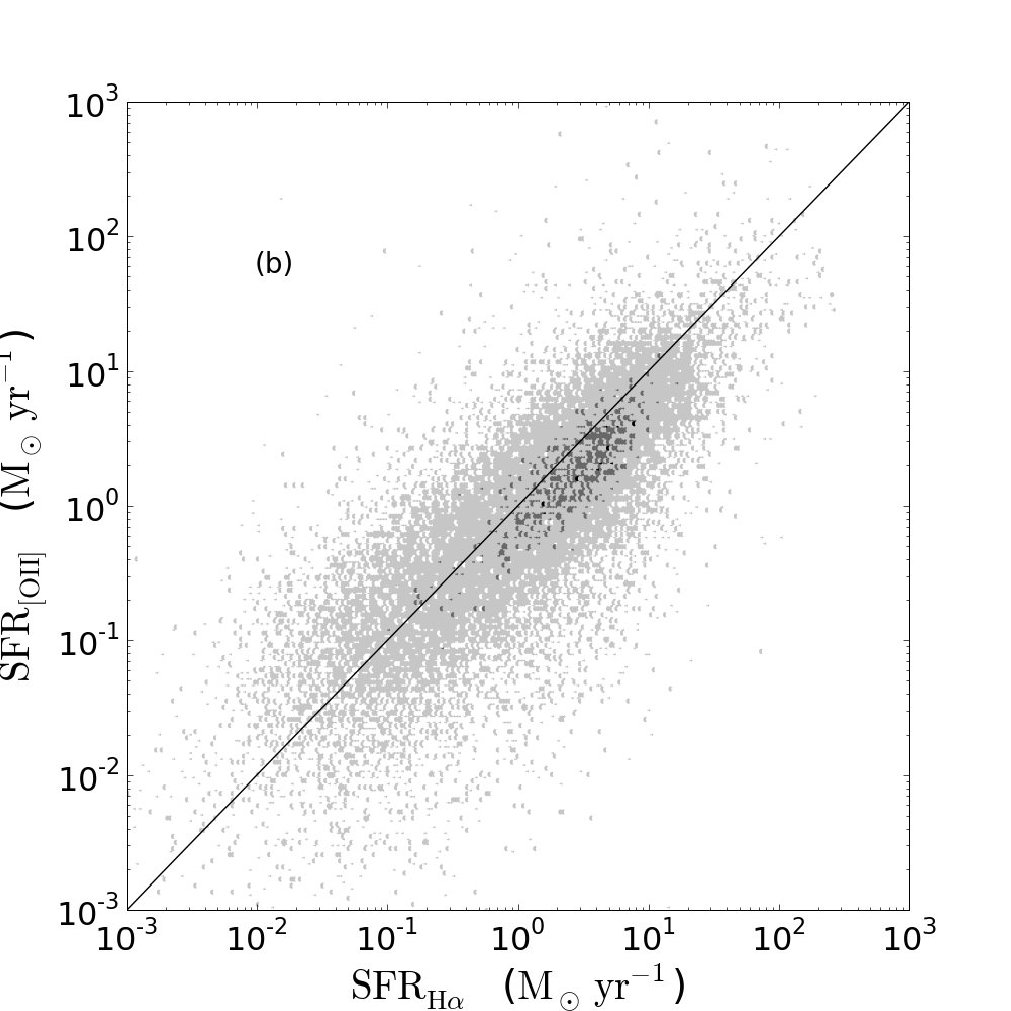}
\includegraphics[width=62mm, height=60mm]{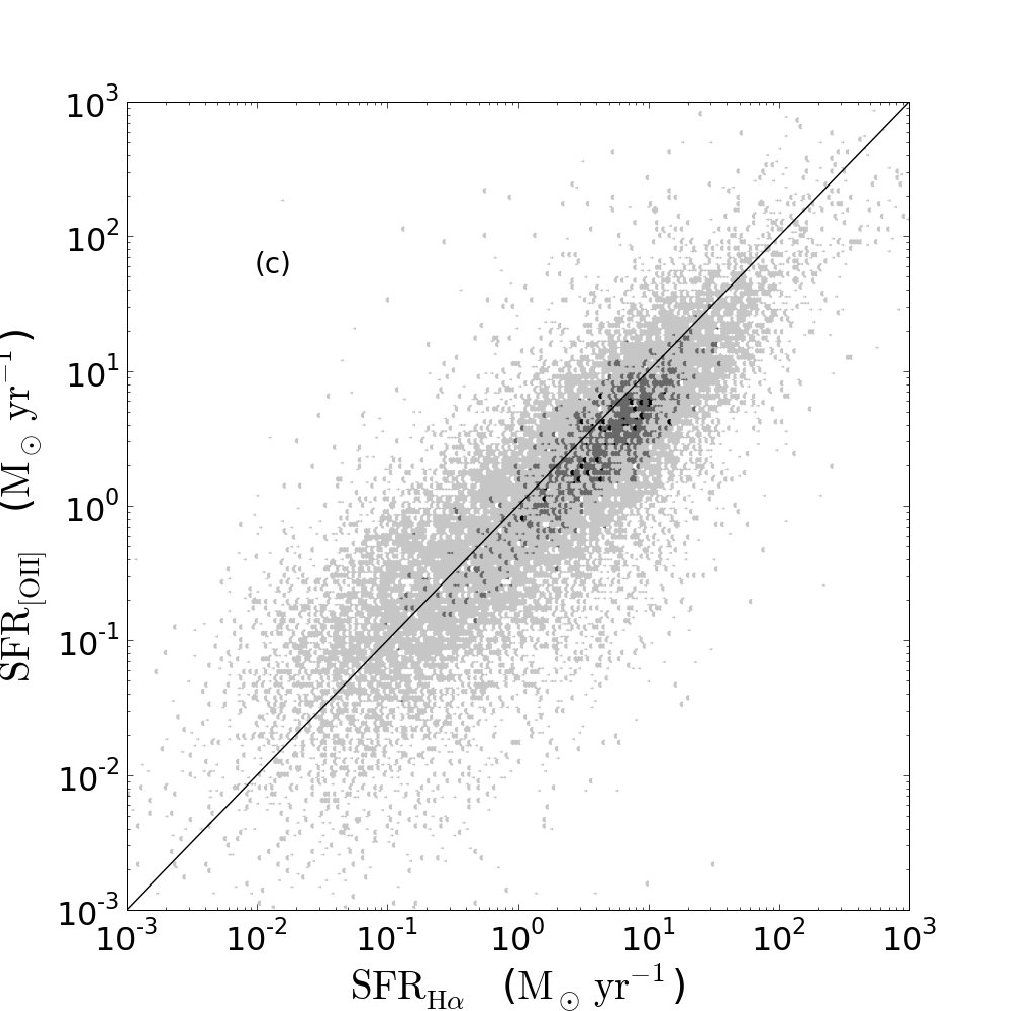}}
\caption{[O{\sc ii}] SFRs as a function of H$\alpha$ SFRs. The [O{\sc ii}] luminosities were corrected based on the extinction at the H$\alpha$ wavelength.
(a) compares the SFR indicators without any extinction corrections. (b) compares SFRs obscuration corrected with the Cardelli et al. (1989) obscuration curve.
(c) compares SFRs obscuration corrected with the FD05 obscuration curve with an $R_{v}$ value of 4.5.}
\label{fig:sfr_ha_fuv_oii}
\end{figure*}

Figures~\ref{fig:sfr_ha_fuv_same_curve} (FUV) and \ref{fig:sfr_ha_nuv_same_curve} (NUV) show the relationship between H$\alpha$, FUV and NUV SFRs after the luminosities have been corrected for
dust obscuration by various obscuration curves. Figure~\ref{fig:sfr_ha_fuv_oii} shows a similar comparison between H$\alpha$ and [O{\sc ii}] SFRs.
The obscuration curves are compared in Figure 5.
There is an overestimation of the UV SFRs when only using the \citet{Cdl:89} curve for both the H$\alpha$ and UV corrections 
(Figures~\ref{fig:sfr_ha_fuv_same_curve}\,a and  \ref{fig:sfr_ha_nuv_same_curve}\,a). The overestimation is larger at higher SFRs.
A similar overestimation is seen when only the \citet{Cal:01} curve is used to correct both H$\alpha$ and UV SFRs 
(Figures~\ref{fig:sfr_ha_fuv_same_curve}\,b and  \ref{fig:sfr_ha_nuv_same_curve}\,b). The overestimation is constant for all SFRs.

Figures~\ref{fig:sfr_ha_fuv_same_curve}\,c and \ref{fig:sfr_ha_nuv_same_curve}\,c show SFRs after the dust corrections were carried out on the appropriate type of emission.
This means that when dust correcting FUV and NUV luminosities a MW obscuration curve \citep{Cdl:89} was used for the nebular
emission parts of UV dust correction and a \citet{Cal:01} curve was used for the continuum parts of UV dust correction as outlined in $\S$3.1. The H$\alpha$ luminosities were corrected using the
same MW obscuration curve \citep{Cdl:89} that was used for the nebular emission correction part of the UV dust corrections. 
There is still an overestimation in the UV and this overestimation is larger than that in panels (a) and (b) in 
Figure~\ref{fig:sfr_ha_fuv_same_curve} and panel (b) in Figure~\ref{fig:sfr_ha_nuv_same_curve}. The overestimation is larger at the high SFR end for both FUV and NUV.

In panels (d), (e) and (f) in Figure~\ref{fig:sfr_ha_fuv_same_curve} the same FD05 curve (i.e. the same $R_{v}$ value) is used for nebular and
continuum parts of UV dust correction and for the dust correction of H$\alpha$ for each panel.
There is good agreement between H$\alpha$ and FUV derived SFRs for an $R_{v}$ value of 4.5 (Figure~\ref{fig:sfr_ha_fuv_same_curve}\,e). 
This value also agrees with the $R_{v}$ value suggested by Calzetti (2001) as well as comparisons made between between FD05 curves and the \citet{Cal:01} curve \citep{FD:03}.
For $R_{v}$ values less than 4.5, the FUV SFRs are overestimated (Figure~\ref{fig:sfr_ha_fuv_same_curve}\,d, $R_{v}=3.5$) and for values above 4.5 
the FUV SFRs become underestimated (Figure~\ref{fig:sfr_ha_fuv_same_curve}\,f, $R_{v}=5.5$),
with respect to the H$\alpha$ SFRs. Both H$\alpha$ and FUV luminosities were corrected using the same FD05 curve.

The corrections applied to the NUV SFR using the FD05 extinction curves only lead to an agreement with H$\alpha$ SFRs at an $R_{v}$ value of 6 (Figure~\ref{fig:sfr_ha_nuv_same_curve}\,f).
Applying an $R_{v}$ value of 4.5 that is consistent with FUV dust corrections overestimates the NUV SFRs, as shown in Figure~\ref{fig:sfr_ha_nuv_same_curve}\,d.
$R_{v}$ values higher than 6 lead to an underestimation of NUV SFRs while lower values lead to an overestimation in the NUV SFRs.
Not only is $R_{v}=6$ inconsistent with the value obtained for the matching between H$\alpha$ and FUV SFRs, it is also too high to be a realistic value \citep{Stn:79,Cdl:89,Cal:01}.
The answer to this dilemma lies in the \citet{Cdl:89} curve. This obscuration curve, 
over-corrects the NUV derived SFRs more than the FUV derived SFRs. A similar result is obtained when the \citet{Stn:79} obscuration curve is used.

In comparing Figures~\ref{fig:sfr_ha_fuv_same_curve} and \ref{fig:sfr_ha_nuv_same_curve} the \citet{Cal:01} obscuration curve (used in panel (b) in both figures) is the only obscuration curve
that does not over-correct NUV more than FUV. This is due to the inclusion of the 2200\,\AA\ feature in the MW and FD05 extinction curves which magnifies the dust corrections. This 
feature is present in all the curves that are discussed here except for the \citet{Cal:01} curve. The sudden increase in the extinction correction near NUV wavelengths is undoubtedly the source of
the overcorrection seen in Figure~\ref{fig:sfr_ha_nuv_same_curve}. For this reason, we calculated the NUV dust correction using an FD05 curve with an $R_{v}$ value of 4.5 but interpolating across 
the 2200 $\textrm{\AA}$ feature (as shown in Table~\ref{table:curve}, where $A_{v}$ is the total attenuation).
The results seen in Figure~\ref{fig:sfr_ha_nuv_same_curve}\,d 
are very promising. The H$\alpha$ and NUV SFRs are very closely matched, however, there appears to be a slight overestimation in the NUV SFRs at all SFRs.
This is a result of the interpolation technique used to recreate this obscuration curve without the 2200\,\AA\ feature.
We find that to correct NUV luminosities for the effects of dust, the role of the 2200\,\AA\ feature will have to be reassessed. The implication is that the 2200\,\AA\ feature is a property of MW
dust grains that does not seem to propagate to the global integrated attenuation properties of galaxies.

\begin{table}
\centering
\caption{Modified FD05 curve ($R_{V}=4.5$) without the 2200\,\AA\ feature}
\begin{tabular}{cc} 
\hline\hline \\
$\lambda {\mu}m$ & $k_{\lambda}$ \\ [0.5ex]
\hline \\
$0.0192$ & $8.133+R_{V}$ \\
$0.1503$ & $6.932+R_{V}$ \\
$0.1216$ & $5.621+R_{V}$ \\ 
$0.1550$ & $4.169+R_{V}$ \\
$0.3000$ & $2.684+R_{V}$ \\
$0.3650$ & $1.830+R_{V}$ \\ 
$0.4400$ & $1.000+R_{V}$ \\ 
$0.5480$ & $0.000+R_{V}$ \\
$0.7200$ & $-1.134+R_{V}$ \\ 
$1.0300$ & $-2.323+R_{V}$ \\
$1.2390$ & $-2.786+R_{V}$ \\
$1.6490$ & $-3.336+R_{V}$ \\ 
$2.1920$ & $-3.714+R_{V}$ \\
$3.5920$ & $-4.179+R_{V}$ \\
$4.7770$ & $-4.130+R_{V}$ \\[1ex]
\hline
\end{tabular}
\label{table:curve}
\end{table}

Figure~\ref{fig:sfr_ha_fuv_oii} shows a good correlation between H$\alpha$ and [O{\sc ii}] derived SFRs. 
Both the FD05 ($R_{v}=4.5$) and \citet{Cdl:89} curves show good agreement between H$\alpha$ and [O{\sc ii}] SFRs as seen in Figure~\ref{fig:sfr_ha_fuv_oii}.
Figure~\ref{fig:sfr_ha_fuv_oii}\,a indicates that [O{\sc ii}] luminosities are obscured more than H$\alpha$ as the
H$\alpha$ SFRs are larger than [O{\sc ii}] before the dust corrections. This is expected as shorter wavelengths are more sensitive to dust obscuration.
We find that the metallicity dependence of the [O{\sc ii}] SFRs does not play a significant role in biasing the SFRs derived from [O{\sc ii}] for this sample although it may contribute to the
observed scatter. It may become significant for the low-SFR systems which begin to deviate from the one-to-one line in Figure~\ref{fig:sfr_ha_fuv_oii} \citep{Kwl:01}.

Finally, we turn to an investigation of the utility of the $\beta$ parameter for making obscuration corrections to UV-derived SFRs. As mentioned above, $\beta$ is demonstrated to be a useful
obscuration parameter only for starburst galaxies (Kong et al. 2004). Our sample consists of galaxies with a broad range of SFRs, but containing very few starbursts. 
The distribution of stellar-mass doubling-times, $t_{d}$, ranges from $10^{8}$ to $10^{13}$ years, with the vast majority ($\approx 97\%$) having $t_{d} > 10^{9}$. The definition of a starburst 
is such that their mass doubling times should be less than  $\approx 1$Gyr, so even the very high SFR systems in our sample (in particular) are not starbursts, as they are also among the most 
massive.

Our results (Figure 6) clearly show FUV SFRs, corrected for obscuration using $\beta$, that are significantly overestimated, by up to two orders of magnitude when
compared to H$\alpha$ derived SFRs corrected using the FD05 curve.
As Kong et al. (2004) suggest, when the $\beta$ parameter is applied to non-starburst galaxies the value of $\beta$ will be overestimated, leading to an overestimation in the dust 
present in the galaxy and consequently the inferred SFR.

Panel (b) of Figure 6 shows FUV derived SFRs corrected using a modified $\beta$ attenuation correction to that of \citet{MHC:99}. The corrected attenuation factor was derived from 
the offset seen in FUV derived SFRs compared to H$\alpha$ SFRs when the attenuation factor of \citet{MHC:99} is used to correct FUV luminosities. We describe the attenuation as
\begin{equation}
A = 3.3 + 1.99\beta.
\end{equation}
The galaxies in Figure 6(b) do appear somewhat more evenly distributed on either side of the one-to-one line. There remains, though, a very large
scatter, with an rms of 0.27.

The conversion from luminosities to SFRs is carried out with the assumption that the IMF is constant and independent of the total SFR. 
Evolving IMFs could have an impact on the way luminosities are converted into SFRs, affecting the trends observed so far, but addressing this is deferred to later publications.

\section{SFR History}
The SFR distributions shown in the previous section include galaxies at various stages of their evolution. By comparing these galaxies with evolutionary synthesis models, using PEGASE.2
\citep{FR:97}, we can develop an understanding of the effects that various galaxy parameters have on their evolution and even explain some of the trends and scatter observed in our data.
The output of the evolutionary synthesis model is normalized to a galaxy of one solar mass. To compare the theoretical models with the observed data
we must scale the theoretical output according to the mass of the galaxies we are analysing.

\begin{figure}
\begin{center}
\includegraphics[width=90mm, height=100mm]{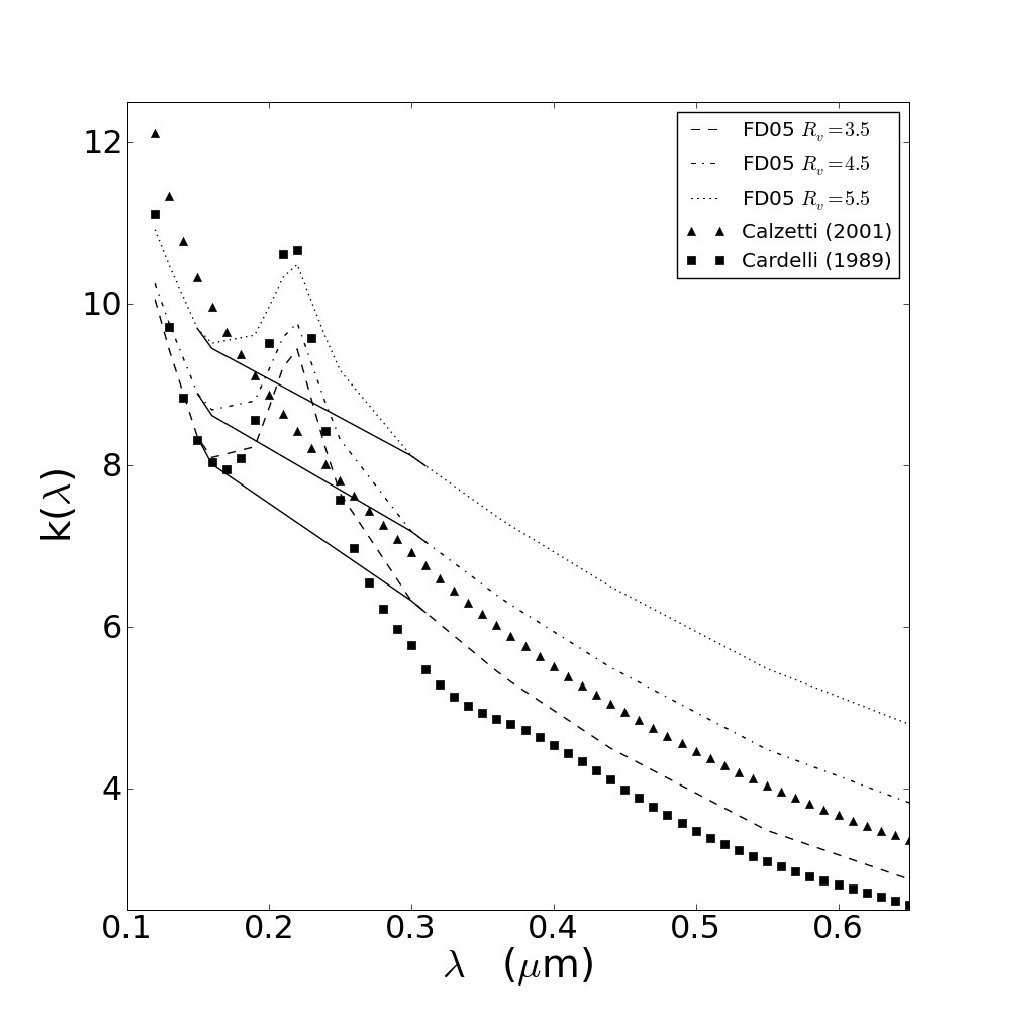}
\caption{The Cardelli et al. (1989), Calzetti (2001) and FD05 ($R_{v} = 3.5, 4.5, 5.5$) obscuration curves. The 3 unbroken lines show the FD05 curves 
with the 2200\,\AA\ feature removed. It must be noted that the Calzetti (2001) curve is the only curve intrinsically without the 2200\,\AA\ feature.}
\end{center}
\label{fig:extinction_curves}
\end{figure}

\begin{figure}
\begin{center}
\includegraphics[width=80mm, height=120mm]{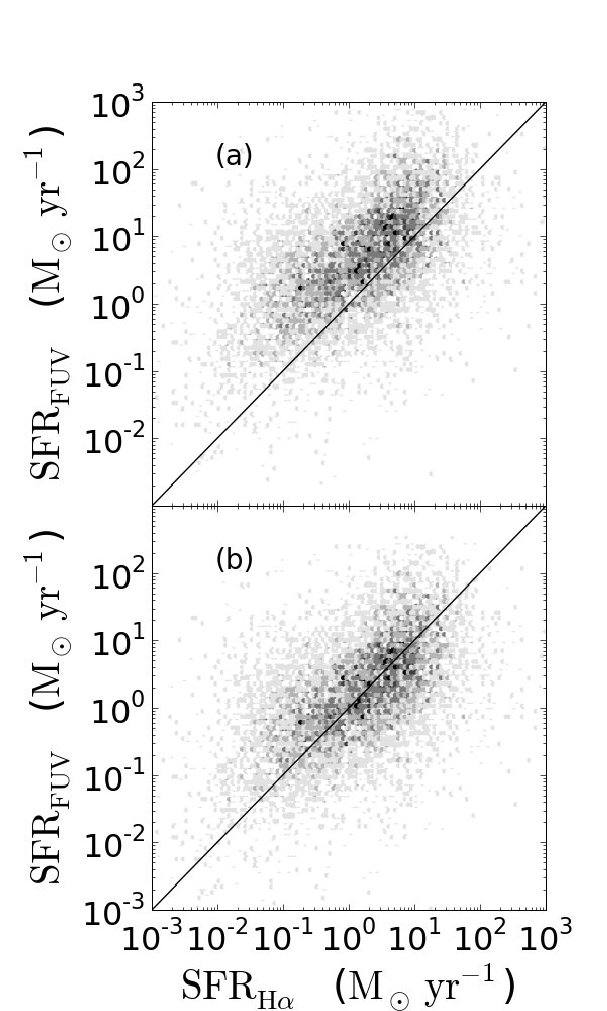}
\caption{The figure compares the H$\alpha$ derived SFR, extinction corrected using a FD05 extinction curve ($R_{v}=4.5$),
against FUV derived SFRs extinction corrected using the $\beta$ parameter. (a) shows the FUV SFR corrected using the attenuation factor in Meurer, Heckman and Calzetti (1999).
(b) is the same as (a) but with FUV SFRs corrected for dust using a modified attenuation factor to that of Meurer, Heckman and Calzetti (1999) chosen to fit the one-to-one line. 
There is a large scatter with an overestimation of the FUV SFRs.}
\end{center}
\label{fig:beta_sfr}
\end{figure}

The galaxy mass is derived using the $g$-band and $i$-band colours and the absolute magnitude in the $i$-band ($M_{i}$).
\begin{equation}
M_{*} = 10^{1.04 + 0.9(g - i) - 0.4M_{i}}, 
\end{equation}
where $g$ and $i$ refer to the SDSS Petrosian magnitudes observed in the $g$ and $i$ bands \citep{Bld:04, Tal:10}.

\begin{figure*}
\centerline{\includegraphics[width=80mm, height=60mm]{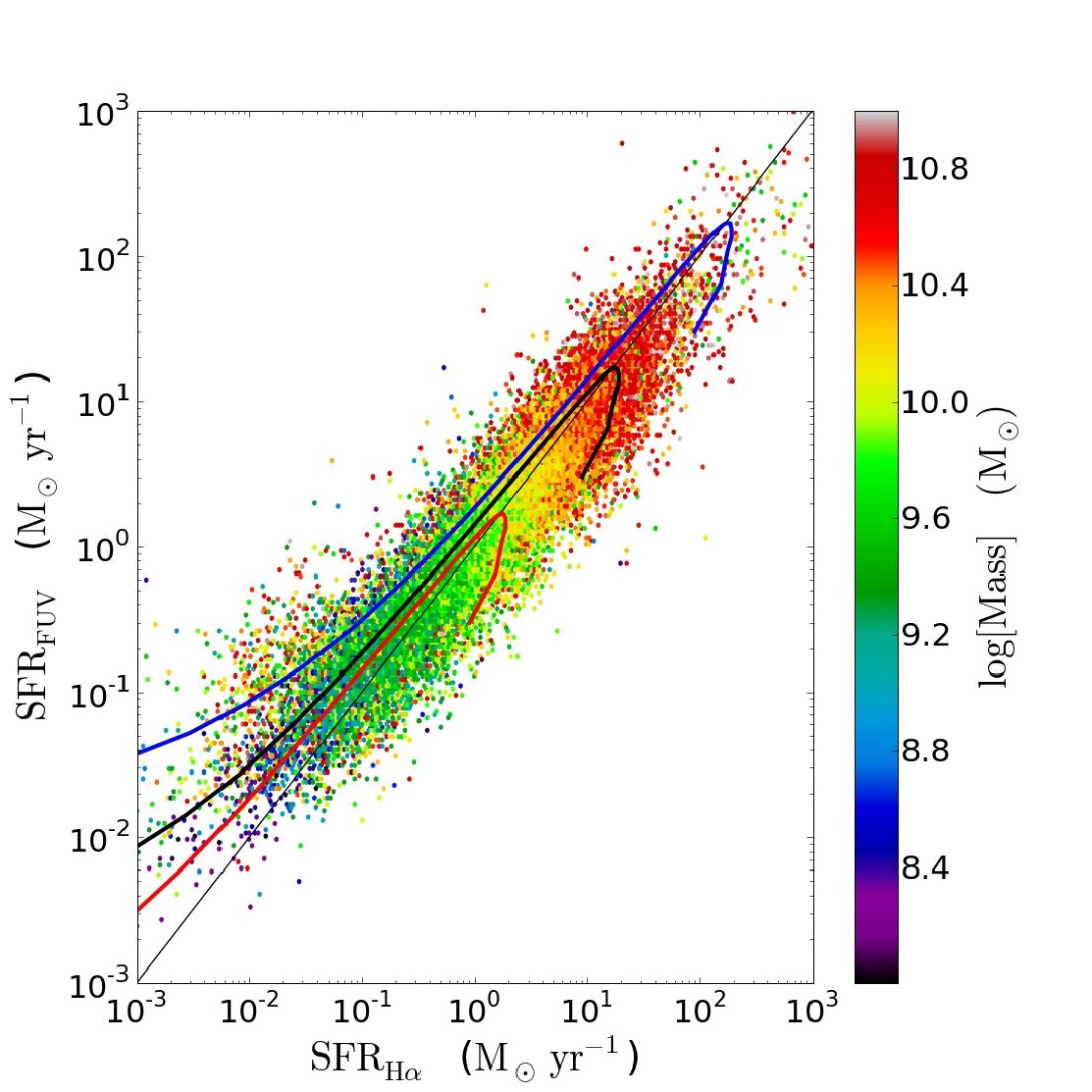}}
\caption{FUV SFRs as a function of H$\alpha$ SFRs compared with PEGASE evolutionary models of galaxies with various masses. The colour scale represents the measured masses of the galaxies.
The regions where the modelled paths agree with the observed masses give an indication of the ages of the galaxies in each part of each curve. The evolutionary paths are for exponentially 
decaying SFRs with masses of $10^{8}\,$M$\sun$ (red), $10^{9}\,$M$\sun$ (black) and $10^{10}\,$M$\sun$ (blue). All paths have a characteristic decay time of 90\,Myr and and 
a \citet{BG:03} IMF.}
\label{fig:sfr_ha_fuv_mass}
\end{figure*}

\begin{figure*}
\centerline{\includegraphics[width=80mm, height=60mm]{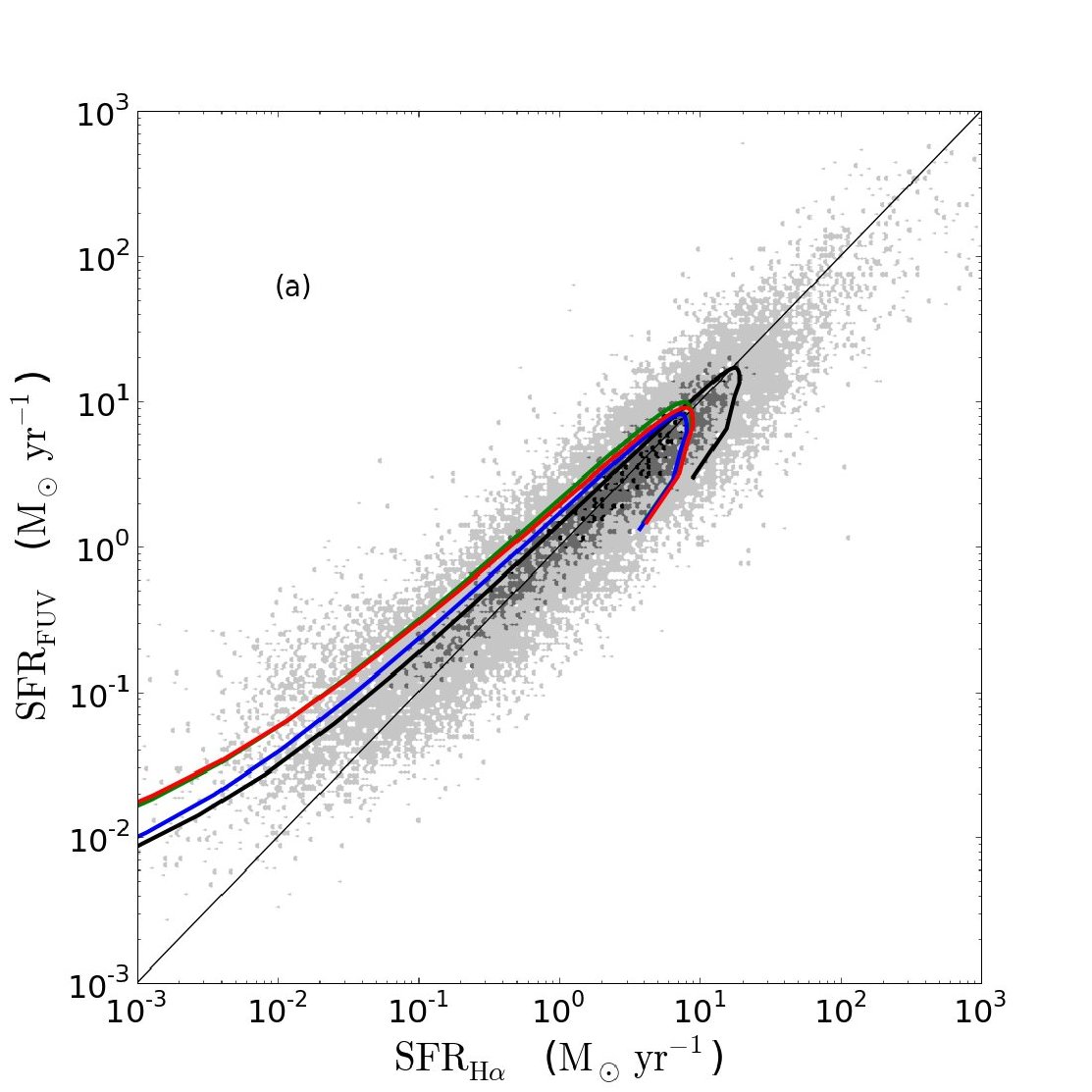}
\includegraphics[width=80mm, height=60mm]{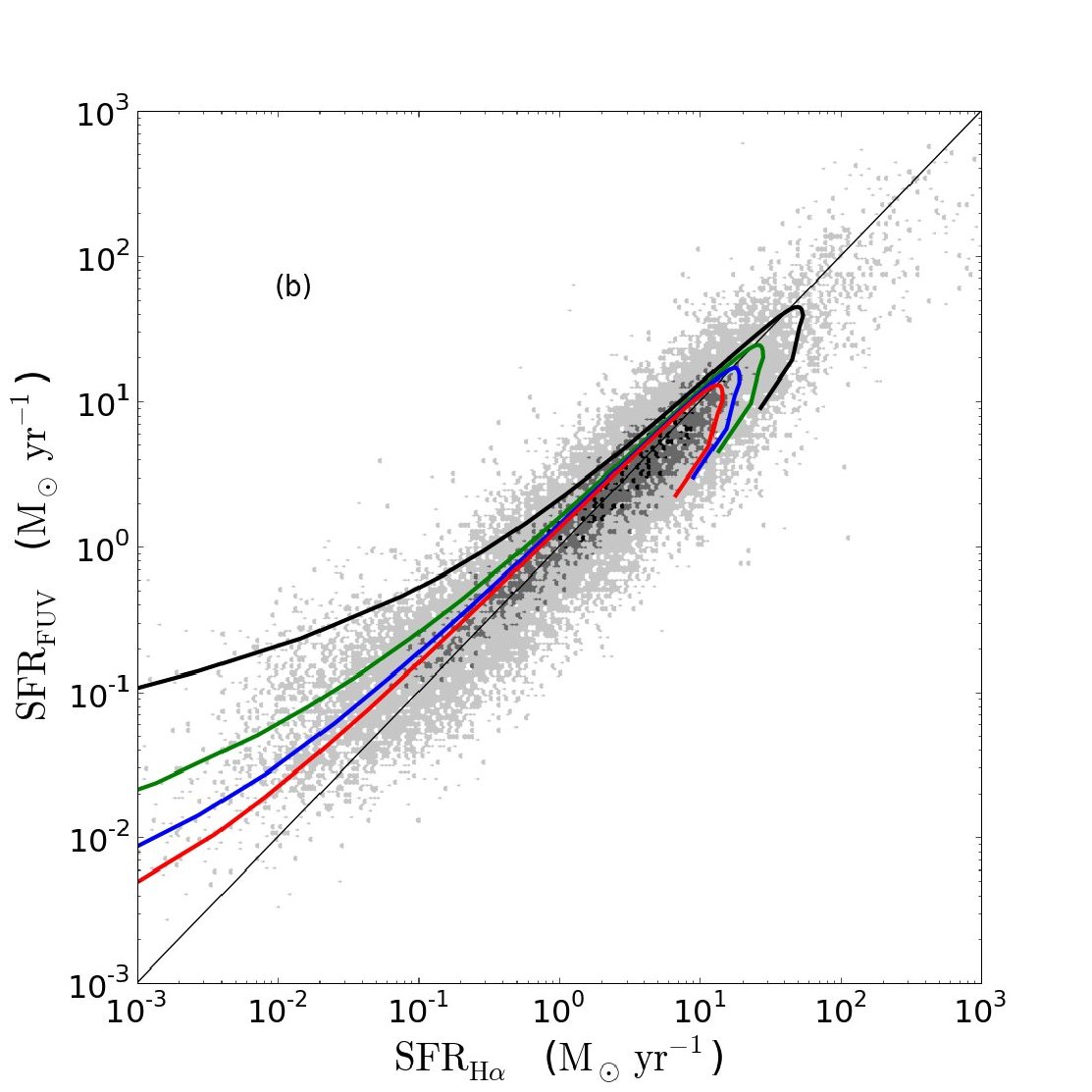}}
\caption{FUV SFRs as a function of H$\alpha$ SFRs compared with PEGASE evolutionary models for different IMFs, panel (a), and characteristic decay times, panel (b).
(a) \citet{Slpt:55} (blue), \citet{Ken:83} (green), \citet{Scl:98} (red) and \citet{BG:03} (black) IMFs are shown.
All IMFs are scaled to a galaxy of mass $10^{9}\,$M$\sun$ and a characteristic decay time of 90\,Myr.
(b) Characteristic decay times of 30 Myr (black), 60 Myr (green), 90 Myr (blue) and 120 Myr (red). All paths were derived using a \citet{BG:03} IMF
scaled to a galaxy of mass $10^{9}\,$M$\sun$.} 
\label{fig:sfr_ha_fuv_IMF_decay_time}
\end{figure*}

Figure~\ref{fig:sfr_ha_fuv_mass} shows how the modeled evolutionary paths compare to the observed measurements. The paths in Figure~\ref{fig:sfr_ha_fuv_mass}
have a \citet{BG:03} IMF and an exponentially declining SFR with $\tau = 90\,$Myr. To give an indication of time on the evolutionary paths, 
the evolutionary paths begin at 0\,Myr and are displayed to $\approx 1000$\,Myr in 
Figure~\ref{fig:sfr_ha_fuv_IMF_decay_time}. The evolutionary paths indicate that the FUV derived
SFRs and H$\alpha$ derived SFRs should strongly agree at all the SFRs that we have obtained. For the case of H$\alpha$ and FUV SFRs this is consistent with our measurements.

For the majority of the galaxies we compare, the masses of observed galaxies and those of the evolutionary paths seem to agree at the ages of 200 to 500\,Myr.
In Figure~\ref{fig:sfr_ha_fuv_IMF_decay_time}, these ages on the evolutionary paths correspond, given the exponentially declining SF histories, to SFRs ranging from 
$10^{1}$ to $10^{-2} M_{\sun}\,yr^{-1}$.
This is consistent across all mass scales in Figure~\ref{fig:sfr_ha_fuv_mass}. The masses of the modelled paths and the observed data do not agree at younger ages indicating
that we do not observe these galaxies at the earliest stages of their lives. Particularly the ``hook'' like path seen at the early stages ($ < 200\,$Myr) is not observed in our data
and does not seem to constitute any of the scatter . At the older end, the masses of the models and the observed galaxies agree up to about 500\,Myr.

Figure~\ref{fig:sfr_ha_fuv_IMF_decay_time}\,a shows that different IMFs can produce different evolutionary paths particularly early in the life of a galaxy.
The \citet{BG:03} IMF begins at a higher SFR and using a \citet{BG:03} IMF will result in a galaxy at a particular age being observed to have a higher SFR 
compared to \citet{Slpt:55}, \citet{Ken:83} or \citet{Scl:98} IMFs. The modeled paths assume a galaxy mass of $10^{9} M_{\sun}$ with a 90\,Myr characteristic decay time.
All the IMFs share similar paths with the \citet{BG:03} IMF showing a better fit to the observed trend than the other 3 IMFs.
The tracks assuming \citet{Slpt:55, Ken:83} and \citet{Scl:98} IMFs follow similar trends at almost the same rate of evolution. 
The evolutionary paths for the IMFs match with the observed values at about the age of 130\,Myrs.

Also of interest is the lower SFR end where the evolutionary paths drift away from the one-to-one line and towards higher FUV-derived SFRs. 
This is clearly visible in the modeled evolutionary tracks. Even though our data does not extend below $10^{-2} M_{\sun}\,yr^{-1}$ there is a mild but definite trend towards higher FUV SFRs 
compared to H$\alpha$ SFRs observed at the low SFR end. 
By varying the characteristic decay times of the evolutionary paths, as in Figure~\ref{fig:sfr_ha_fuv_IMF_decay_time}\,b, various turn-off points can be modeled.
A characteristic decay time of 90Myr seems to agree well with the observed values.
Possible causes for this behaviour will be discussed in \S\,6.

Figures~\ref{fig:sfr_ha_fuv_mass} and \ref{fig:sfr_ha_fuv_IMF_decay_time} show good model agreement with the observed SFR distributions.
It is clear that the evolution of galaxies with $ 10^{-2} M_{\sun}\,yr^{-1} < SFR < 10^{2} M_{\sun}\,yr^{-1}$ can be modeled effectively.

\section{Discussion}

\subsection{Obscuration curves}

We have presented an analysis of the dust corrections in a sample of 31058 galaxies from the GAMA survey.
We analysed several techniques used to correct for the dust obscuration in galaxies. 

The Calzetti formalism using the Calzetti (2001) obscuration curve
has been deemed to be one of the best for applying obscuration corrections to FUV emission in starburst galaxies but we find that in our sample of non-starburst galaxies, it overestimates the UV SFR.
We find that a theoretically modelled obscuration mechanism by FD05 provides SFRs well-matched between FUV and H$\alpha$.
In addition, we use a Balmer decrement to derive the colour excess suffered by the stellar continuum, $E(B-V)_{star}$, to complete our UV dust corrections.

The effect of applying dust corrections is shown in Figure~\ref{fig:sfr_ha_corr_uncorr_fuv_uncorr} where the large increase in SFRs after applying dust corrections 
(to the H$\alpha$ emission in this case) is a consequence of high star formation rate galaxies having much higher dust obscuration levels 
\citep{Cal:01, Hpk:01, Hpk:03, Afn:03, vdS:07}.
The same figure illustrates the lack of dust in low star forming systems as galaxies at the very low SFR end maintain similar SFRs after dust corrections.

We explored several obscuration curves used to correct emission line luminosities,
such as the \citet{Stn:79} and \citet{Cdl:89} MW obscuration curves as well the FD05 curves. It is important that the appropriate type of obscuration curve is applied for different types
of emission. The FUV emission from galaxies contain contributions from older and lower mass stellar populations \citep{Bel:03} which would have gradually moved away from the clouds of 
gas that formed them, whereas H$\alpha$ and [O{\sc ii}] lines are emitted from the [H{\sc ii}] regions ionised by hot young stars that are contained within those clouds of gas. 
For this reason, different absorption curves could potentially apply to stellar and nebular emission, highlighting the importance of using the correct obscuration curve for each type of emission. 

This is clearly illustrated in Figures~\ref{fig:sfr_ha_fuv_same_curve}\,a and \ref{fig:sfr_ha_nuv_same_curve}\,a where using the \citet{Cdl:89} curve for both H$\alpha$ and UV 
overestimate the UV SFRs. The \citet{Cdl:89} curve is appropriate for correcting nebular emission, and is more likely to over-correct for dust when applied to the UV stellar 
continuum.

The overestimation of the UV SFRs when the \citet{Cal:01} curve is used (Figures~\ref{fig:sfr_ha_fuv_same_curve}\,b and \ref{fig:sfr_ha_fuv_same_curve}\,c 
and \ref{fig:sfr_ha_nuv_same_curve}\,b and \ref{fig:sfr_ha_nuv_same_curve}\,c
can have several explanations. As highlighted by Calzetti, \citep{Cal:01}, this curve is only applicable to starburst galaxies. 
It was also based on a small sample of 37 starburst galaxies using a 
universal extinction as opposed to a galaxy specific approach. It must be noted that a starburst galaxy is not the same as a high SFR galaxy.
A starburst galaxy requires a short mass-doubling time compared to a non-starburst galaxy reflecting an atypically high SFR.
Most of the galaxies in our sample have mass doubling times consistent with not being in starburst mode.
For this reason, it is conceivable that the \citet{Cal:01} obscuration corrections would overestimate the UV derived SFR of the galaxies in our sample.
The processes that lead to starbursts such as interaction of galaxies, tidal shear in the solid body rotation and secular evolution of bars 
\citep{LT:78, Snd:88, Ken:90, NHS:96, KCY:93} would not necessarily occur in non-starburst galaxies and hence the spectral output can be different to a starburst galaxy.

The FD05 extinction curve, with an $R_{v}$ value of 4.5 is successful in providing matching H$\alpha$, FUV and [O{\sc ii}] derived SFRs at all SFRs that we have tested 
($10^{-2} M_{\sun}\,yr^{-1} < SFR < 10^{2} M_{\sun}\,yr^{-1}$). An $R_{v}$ value of 4.5 agrees with values prescribed by \citet{Cal:01}. 
The FD05 curves are theoretical models based on the physical density of the turbulent structure and hence do not carry
with them any of the constraints that other curves have in their application. As the model deals only with the absorption and has nothing to do with the properties of the 
emission sources, the FD05 curves can be applied to any given emission as they are not situation dependent.

In the NUV, using the same FD05 curve as for the FUV ($R_{v}=4.5$) leads to an overestimation in the NUV SFRs compared to H$\alpha$ derived SFRs.
There is agreement when $R_{v}=6$ but this is inconsistent with other measured values, and with the value of $R_{v}$ found to reconcile the FUV, H$\alpha$ and [O{\sc ii}] measurements in our 
analysis. The fact that the NUV SFRs are overestimated more than the FUV SFRs in all MW obscuration curves and not in the Calzetti
(2001) curve suggest that this effect is due to the 2200\,\AA\ feature being included in the extinction curve. The NUV being the only SFR indicator that otherwise does not 
agree with the other indicators lends weight to this argument as none of those wavelengths fall within the range of the 2200\,\AA\ feature. We show that the H$\alpha$ and 
NUV derived SFRs agree very well if the 2200\,\AA\ feature is eliminated in an FD05 curve with an $R_{v}$ value of 4.5. This provides us with a consistent extinction
correction curve for all the SFR indicators. The role of the 2200\,\AA\ feature will require further analysis in future work.

The use of the $\beta$ parameter in estimating obscuration corrections for our sample of largely quiescent galaxies shows that intrinsic SFRs will be overestimated. 
Kong et al. (2004) attribute this to a secondary effect based on SFR histories.

\subsection{SFR Histories}

We applied galaxy evolutionary synthesis models to the comparison between H$\alpha$ and FUV derived SFRs, and there was good agreement for all SFRs. The evolutionary paths were
compared according to galaxy mass, characteristic decay time and IMF (assuming an exponentially decaying SFR).

We showed that by comparing the evolutionary paths for varying galaxy masses with observed data, we can identify a range of ages that the observed galaxies would fall into. 
This range was estimated to be between 200\,Myr and 500\,Myr. There is a continuous variation in the ages making it 
difficult to identify any effects on the ages of galaxies due to mass. The IMFs we have tested show similar trends with a \citet{BG:03} IMF showing the best agreement with the observed data
(Figure~\ref{fig:sfr_ha_fuv_IMF_decay_time}\,a). This is particularly clear at the low SFR end with the \citet{BG:03} IMF following the observed SFR trend better than the other 3 IMFs.
The \citet{BG:03} IMF has a faster evolution, where at any given SFR, the \citet{BG:03} evolutionary path has an age that is about 50Myr higher than the other tested IMFs.

The characteristic decay time was used to show how galaxies behave as they age: At low SFRs, a turn-off is 
expected away from the one-to-one line, towards the FUV SFR axis, as the H$\alpha$ derived SFRs decrease faster than FUV derived SFRs.
Our data does not extend to such low SFRs, but using PEGASE models we can recreate a predicted turn-off had our data extended to such low SFRs. The turn-off point
is highly dependent on the characteristic decay time of SFRs with higher characterised decay times causing the turn off to occur later.

One of the proposed mechanisms for this behaviour is the
Integrated Galaxy IMF (IGIMF) theory \citep{Pfl:07, Pfl:09}. \citet{Pfl:07, Pfl:09} constructed an integrated galactic initial mass function by combining all
IMFs of all young star clusters. Using the IGIMF they predicted that the H$\alpha$ luminosity of galaxies should decrease faster than UV luminosities at low SFRs 
as the H$\alpha$ luminosity of star-forming dwarf galaxies decreases faster with decreasing SFR than the UV luminosity at low SFRs. This is clearly observed in the evolutionary paths in
Figures~\ref{fig:sfr_ha_fuv_IMF_decay_time}\,a and b. \citet{Pfl:07, Pfl:09} predict this turn-off to occur at around $10^{-2} M_{\sun}\,yr^{-1}$. As seen in
Figure~\ref{fig:sfr_ha_fuv_IMF_decay_time}\,b this turn-off can be replicated by a 90 Myr characteristic decay time which also agrees well with the observed data at higher SFRs.
Even though our data does not extend far below an SFR of $10^{-2} M_{\sun}\,yr^{-1}$, the beginnings of such a trend can be
observed in Figures~\ref{fig:sfr_ha_fuv_same_curve}\,e, \ref{fig:sfr_ha_fuv_mass} and \ref{fig:sfr_ha_fuv_IMF_decay_time}.



\section{Conclusion}
We find that an FD05 obscuration curve with an $R_{v}$ value of 4.5 gives the best agreement across all SFR indicators provided that the 2200\,\AA feature is not included.
These SFR distributions agree well with modeled evolutionary paths.
We also show that the $\beta$ parameter is not suitable for use with GALEX data. 

Our main objective has been to obtain accurate dust extinction corrections for H$\alpha$ and FUV and present a clear and concise formalism on how this is to be done based on various
obscuration laws described in the literature. While the dust corrections were largely successful, we have identified key areas for further research, particularly in the 
significance of the 2200\,\AA\ in obscuration corrections. Analysing the effectiveness of extinction curves, particularly with IR data will be of great significance for future 
work.

\section*{Acknowledgements}
D.B.W acknowledges the support provided by the Denison Scholarship from the School of Physics,
University of Sydney. A.M.H. acknowledges support provided by the Australian Research Council through a QEII Fellowship (DP0557850).

\label{lastpage}

\end{document}